\documentclass[prodmode,acmtodaes]{acmsmall} 

\newcommand{\hidetodos}{false}

\usepackage{pgf,tikz,pgfplots,booktabs,ifthen,pgfplotstable,etoolbox,color,xspace}
\usetikzlibrary{arrows,automata,backgrounds,calc,chains,decorations,matrix,positioning,patterns,petri,shadows,shapes,decorations.pathmorphing,mindmap,fit}
\usetikzlibrary{circuits.ee.IEC}
\usetikzlibrary{pgfplots.groupplots}
\usepackage{algorithm,algpseudocode}
\usepackage{amssymb,makeidx,epic,eepic,latexsym,url,graphicx,amsmath}
\usepackage{textcomp}
\usepackage{cleveref}
\usepackage{graphicx}
\usepackage{rotating}
\usepackage{todonotes}
\usepackage{enumerate}

\pgfplotsset{compat=1.12}

\usepackage[binary-units]{siunitx}
\usepgfplotslibrary{groupplots}
\usepackage{microtype}

\usepackage{listings,multicol}
\ifthenelse{\equal{\hidetodos}{true}}
{
	\newcommand{\todophilipp}[1]{}
	\newcommand{\todoandrew}[1]{}
	\newcommand{\todosebastian}[1]{}
	\newcommand{\numberofpages}[1]{}
}{
	\newcommand{\todophilipp}[1]{\todo[inline,color=red!80,caption={}]{Philipp: #1}}
	\newcommand{\todoandrew}[1]{\todo[inline,color=green!80,caption={}]{Andrew: #1}}
	\newcommand{\todosebastian}[1]{\todo[inline,color=blue!80]{Sebastian: #1}}
	\newcommand{\numberofpages}[1]{\todo[inline,color=yellow!80]{Number of pages: #1}}
}

\Crefname{lstlisting}{Listing}{Listings}

\newcommand{\minisection}[1]{\noindent {\textbf{#1.}}}
\usepackage[toc, nonumberlist]{glossaries}

\newacronym{ISO}{ISO}{International Organization for Standardization}
\newacronym{CSV}{CSV}{Comma-Separated Value}
\newacronym{AVB}{AVB}{Audio Video Bridging}
\newacronym{XML}{XML}{Extensible Markup Language}
\newacronym{FIBEX}{FIBEX}{Field Bus Exchange Format}
\newacronym{XSD}{XSD}{XML Schema Definition}
\newacronym{JAXB}{JAXB}{Java Architecture for XML Binding}
\newacronym{ECU}{ECU}{Electronic Control Unit}
\newacronym{DOM}{DOM}{Document Object Model}
\newacronym{SAX}{SAX}{Simple API for XML}
\newacronym{JRE}{JRE}{Java Runtime Environment}
\newacronym{PDU}{PDU}{Protocol Data Unit}
\newacronym{CAN}{CAN}{Controller Area Network}
\newacronym{LIN}{LIN}{Local Interconnect Network}
\newacronym{MOST}{MOST}{Media Oriented Systems Transport}
\newacronym{NIT}{NIT}{Network Idle Time}
\newacronym{TDMA}{TDMA}{Time Division Multiple Access}
\newacronym{TUM}{TUM}{Technische Universität München}
\newacronym{CREATE}{CREATE}{Campus for Research Excellence And Technological Enterprise}
\newacronym{ASIL}{ASIL}{Automotive Safety Integrity Levels}
\newacronym{EDF}{EDF}{Earliest Deadline First}
\newacronym{SDF}{SDF}{Shortest Deadline First}
\newacronym{CSMA}{CSMA}{Carrier Sense Multiple Access}
\newacronym{CSMACD}{CSMA/CD}{\gls{CSMA} Collision Detection}
\newacronym{CSMACR}{CSMA/CR}{\gls{CSMA} Collision Resolution}
\newacronym{CSMACA}{CSMA/CA}{\gls{CSMA} Collision Avoidance}
\newacronym{IEEE}{IEEE}{Institute of Electrical and Electronics Engineers}
\newacronym{IP}{IP}{Internet Protocol}\newacronym{TTP}{TTP}{Time-Triggered Protocol}
\newacronym{TTA}{TTA}{Time-Triggered Architecture}
\newacronym{ILP}{ILP}{Integer Linear Program}
\newacronym{NIP}{NIP}{Non-linear Integer Program}
\newacronym{MIP}{MIP}{Mixed Integer Program}
\newacronym{TCP}{TCP}{Transmission Control Protocol}
\newacronym{VLAN}{VLAN}{Virtual Local Area Network}
\newacronym{QoS}{QoS}{Quality of Service}
\newacronym{SRP}{SRP}{Stream Reservation Protocol}
\newacronym{EMI}{EMI}{Electromagnetic Interference}
\newacronym{PHY}{PHY}{physical layer}
\newacronym{API}{API}{Application Programming Interface}
\newacronym{AUTOSAR}{AUTOSAR}{AUTomotive Open System ARchitecture}
\newacronym{ADAS}{ADAS}{Advanced Driver Assistance System}
\newacronym{WCRT}{WCRT}{Worst Case Response Time}
\newacronym{FTDMA}{FTDMA}{Flexible Time Division Multiple Access}
\newacronym{FCFS}{FCFS}{First Come First Serve}
\newacronym{RR}{RR}{Round Robin}
\newacronym{SMF}{SMF}{Shortest Message First}
\newacronym{ABS}{ABS}{Anti-lock Braking System}
\newacronym{TSN}{TSN}{Time-Sensitive Networking}
\newacronym{RTC}{RTC}{Real-Time Calculus}
\newacronym{ASAM}{ASAM}{Association for Standardisation of Automation and Measuring Systems}
\newacronym{OSI}{OSI}{Open Systems Interconnect}
\newacronym{UDP}{UDP}{User Datagram Protocol}
\newacronym{TTL}{TTL}{Time To Live}
\newacronym{MAC}{MAC}{Message Authentication Code}
\newacronym{PCP}{PCP}{Priority Code Point}
\newacronym{DEI}{DEI}{Drop Eligibility Indicator}
\newacronym{VID}{VID}{VLAN Identifier}
\newacronym{TPID}{TPID}{Tag Protocol Identifier}
\newacronym{TCI}{TCI}{Tag Control Information}
\newacronym{IANA}{IANA}{Internet Assigned Numbers Authority}
\newacronym{PKINIT}{PKINIT}{Public Key Cryptography for Initial Authentication in Kerberos}
\newacronym{OEM}{OEM}{Original Equipment Manufacturer}
\newacronym[\glslongpluralkey={Certificate Authorities}]{CA}{CA}{Certificate Authority}
\newacronym{SSL}{SSL}{Secure Sockets Layer}
\newacronym{TLS}{TLS}{Transport Layer Security}
\newacronym{KDC}{KDC}{Key Distribution Center}
\newacronym{TGS}{TGS}{Ticket Granting Server}
\newacronym{TGT}{TGT}{Ticket Granting Ticket}
\newacronym{RNG}{RNG}{Random Number Generator}
\newacronym{PUF}{PUF}{Physically Unclonable Function}
\newacronym{CRL}{CRL}{Certificate Revocation List}
\newacronym{AS}{AS}{Authentication Server}
\newacronym{AES}{AES}{Advanced Encryption Standard}
\newacronym{SHE}{SHE}{Secure Hardware Extension}
\newacronym{HIS}{HIS}{Hersteller Initiative Software}
\newacronym{CMAC}{CMAC}{Cipher-based Message Authentication Code}
\newacronym{ECB}{ECB}{Electronic Code Block}
\newacronym{CBC}{CBC}{Cipher Block Chaining}
\newacronym{CTR}{CTR}{Counter}
\newacronym{OBD}{OBD}{On-Board Diagnosis}
\newacronym{CRC}{CRC}{Cyclic Redundancy Check}
\newacronym{HSM}{HSM}{Hardware Security Module}
\newacronym{EU}{EU}{European Union}
\newacronym{SeVeCom}{SeVeCom}{Secure Vehicle Communication}
\newacronym{TESLA}{TESLA}{Timed Efficient Stream Loss-Tolerant Authentication}
\newacronym{MILP}{MILP}{Mixed-Integer Linear Program}
\newacronym{PRECIOSA}{PRECIOSA}{Privacy Enabled Capability In Co-Operative Systems and Safety Applications}
\newacronym{OTA}{OTA}{Over-The-Air}
\newacronym{IDS}{IDS}{Intrusion Detection System}
\newacronym{TPM}{TPM}{Trusted Platform Module}
\newacronym{OCSP}{OCSP}{Online Certificate Status Protocol}
\newacronym{FD}{FD}{Flexible Data-Rate}
\newacronym{ACL}{ACL}{Access Control List}
\newacronym{TCG}{TCG}{Trusted Computing Group}
\newacronym{NIST}{NIST}{National Institute of Standards and Technology}
\newacronym{ECDSA}{ECDSA}{Elliptic Curve Digital Signature Algorithm}
\newacronym{CCM}{CCM}{Counter with CBC-MAC}
\newacronym{DES}{DES}{Data Encryption Standard}
\newacronym{SHA}{SHA}{Secure Hash Algorithm}
\newacronym{ECC}{ECC}{Elliptic Curve Cryptography}
\newacronym{M-MAC}{M-MAC}{Mixed Message Authentication Code}
\newacronym{MAD}{MAD}{Median Absolute Deviation}
\newacronym{RFC}{RFC}{Request For Comments}
\newacronym{GUI}{GUI}{Graphical User Interface}

\acmVolume{22}
\acmNumber{2}
\acmArticle{25}
\acmYear{2017}
\acmMonth{3}
\doi{http://dx.doi.org/10.1145/2960407}

\makeatletter
\def\@copyrightpermission{\relax}
\makeatother

\begin{document}

\newcommand{\papertitle}{Security in Automotive Networks:\\Lightweight Authentication and Authorization}

\markboth{P. Mundhenk et al.}{\papertitle}


\title{\papertitle}
\author{PHILIPP MUNDHENK
\affil{TUM CREATE Limited, Singapore}
ANDREW PAVERD
\affil{Aalto University, Finland}
ARTUR MROWCA
\affil{TUM CREATE Limited, Singapore}
SEBASTIAN STEINHORST
\affil{TUM CREATE Limited, Singapore}
MARTIN LUKASIEWYCZ
\affil{TUM CREATE Limited, Singapore}
SUHAIB A. FAHMY
\affil{University of Warwick, United Kingdom}
SAMARJIT CHAKRABORTY
\affil{Technische Universit{\"a}t M{\"u}nchen, Germany}}

\begin{abstract}
With the increasing amount of interconnections between vehicles, the attack surface of internal vehicle networks is rising steeply.
Although these networks are shielded against external attacks, they often do not have any internal security to protect against malicious components or adversaries who can breach the network perimeter.
To secure the in-vehicle network, all communicating components must be authenticated, and only authorized components should be allowed to send and receive messages.
This is achieved through the use of an authentication framework.
Cryptography is widely used to authenticate communicating parties and provide secure communication channels (e.g. Internet communication).
However, the real-time performance requirements of in-vehicle networks restrict the types of cryptographic algorithms and protocols that may be used.
In particular, asymmetric cryptography is computationally infeasible during vehicle operation.

In this work, we address the challenges of designing authentication protocols for automotive systems.
We present Lightweight Authentication for Secure Automotive Networks (LASAN), a full life-cycle authentication approach.
We describe the core LASAN protocols and show how they protect the internal vehicle network while complying with the real-time constraints and low computational resources of this domain.
By leveraging on the fixed structure of automotive networks, we minimize bandwidth and computation requirements.
Unlike previous work, we also explain how this framework can be integrated into all aspects of the automotive product life cycle, including manufacturing, vehicle maintenance and software updates.
We evaluate LASAN in two different ways:
Firstly, we analyze the security properties of the protocols using established protocol verification techniques based on formal methods.
Secondly, we evaluate the timing requirements of LASAN and compare these to other frameworks using a new highly modular discrete event simulator for in-vehicle networks, which we have developed for this evaluation.
\end{abstract}




\acmformat{Philipp Mundhenk, Andrew Paverd, Artur Mrowca, Sebastian Steinhorst, Martin Lukasiewycz, Suhaib A. Fahmy, and Samarjit Chakraborty, 2017. \papertitle.}

\begin{bottomstuff}
This work was financially supported by the Singapore National Research Foundation under its Campus for Research Excellence And Technological Enterprise (CREATE) programme.

Authors' addresses: P. Mundhenk, A. Mrowca, S. Steinhorst, and M. Lukasiewycz, TUM CREATE, 1
Create Way, \#10-02, Singapore 138602, Singapore; emails: \{philipp.mundhenk, artur.mrowca, sebastian.
steinhorst, martin.lukasiewycz\}@tum-create.edu.sg; A. Paverd, Department of Computer Science, Aalto
University, Konemiehentie 2, Espoo, FI-02150, Finland; email: andrew.paverd@ieee.org; S. A. Fahmy,
School of Engineering, Library Road, University of Warwick, Coventry, CV4 7AL, United Kingdom; email:
s.fahmy@warwick.ac.uk; S. Chakraborty, Institute for Real-Time Computer Systems, Technical University
of Munich, Arcisstr. 21, 80290 Munchen, Germany; email: samarjit@tum.de.
\end{bottomstuff}

\maketitle

\section{Introduction}
\label{sec:Introduction}
%
The rapidly increasing connectedness of modern cars leads to new challenges in the security of inter- and intra-vehicle communication.
As increasing numbers of vehicles are being connected to the outside world, the exposure risk of safety-critical systems rises significantly.
With the multitude of communication interfaces, it is very difficult, or even impossible, to reliably control all entry points into the vehicle or shield the vehicular network with firewalls.
Hence, it is important that, besides the external access points, communication within a vehicle is secured.
This does not only hold if the external protections are breached, but also for targeted attacks, e.g., via the internal \gls{OBD} port, the infotainment system or the telematics unit.
With the introduction of networked comfort and entertainment functions, as well as \glspl{ADAS}, vehicles are more readily connected to external networks, such as car-to-x networks and the Internet.
This trend towards interconnectivity continues in the vehicle interior.
Increasingly, passengers and drivers connect smartphones and other mobile devices to the vehicle and perform comfort and control functions, such as music streaming, phone access, but also vehicle related functions, such as control of air-conditioning and the vehicle locks.
This increase in interconnections also raises the attack surface of the vehicle.
This holds especially for the integration of vehicles in car-to-x networks.
In these networks, safety-critical data is transmitted, allowing deep insights into the state of a car and high influence on other cars.
The impact of a malicious attack on automotive safety-critical systems can be devastating, including high financial damages and even loss of life.
To mitigate the effects of such attacks, protection mechanisms have been developed to limit unauthorized access to in-vehicle communication and minimize the number of attack vectors \cite{mundhenk2015a,Groza2012,van2011canauth,7030108}.
To allow real-time behavior, these protection mechanisms are based on symmetric cryptography, requiring pre-shared keys.
Pre-programming these keys opens another attack vector, especially if the same keys are used throughout the lifetime of a vehicle and across multiple vehicles.
Thus, it is necessary to generate and exchange keys at runtime.
Only by using dynamic keys and limiting the duration of key validity, can secure in-vehicle communication be provided.


\minisection{Security}
It is to be noted that no security mechanism can be considered 100\% secure.
Security is highly dependent on the capabilities (computational, knowledge) of an attacker, which typically develop over time.
However, the security of a concept, such as a protocol, at the time of development can be formally verified.
This ensures that the protocol is free of flaws that allow circumvention.
Formal verification of the proposed protocols is presented in this work.

The level of security of any measure further depends on the algorithms and parameters chosen for protection.
The algorithms are typically either of symmetric (with pre-shared key) or asymmetric (without pre-shared key) type.
The parameters depend on the algorithms and include for almost all algorithms the length of the encryption key, as well as for some algorithms the chosen encryption basis, e.g., the exponent in case of RSA or the parameters for the selected curve in \gls{ECC}.

Typically, security is analyzed across the metrics of Confidentiality, Integrity, and Availability.
In terms of network communication, Confidentiality describes the security of a message against being read by an attacker.
Integrity describes protection against creation or alteration of messages, and Availability describes the security against interruption of messages.
In this work, we seek to enable the protection of all three aspects through a secure key exchange, allowing the distribution of keys to \glspl{ECU}, which in turn are enabled to encrypt their messages, thus adhering to Confidentiality and Integrity.
At the same time, the authentication of \glspl{ECU} and authorization of message streams helps to ensure that only legitimate \glspl{ECU} can participate in the communication and perpetrators are easier to detect, thus enabling simplified enforcement of Availability.

\begin{figure}
	\centering
	\includegraphics[width=\hsize]{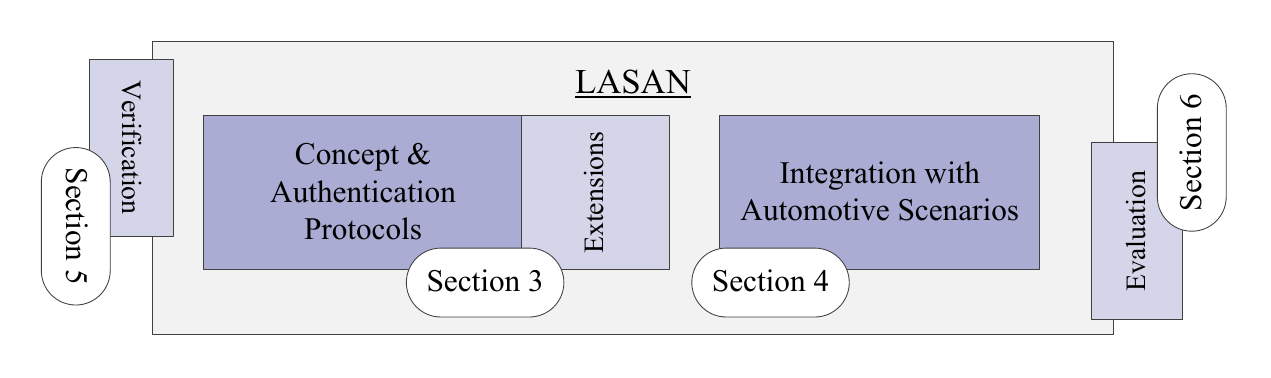}
	\caption{The Lightweight Authentication for Secure Automotive Networks (LASAN) consists of the concept of the authentication protocols to ensure real-time behavior and full security, as well as the protocols and processes required to integrate these authentication protocols into automotive scenarios. We verify and evaluate LASAN for security and performance to show its superior qualities over existing protocols.}
	\label{fig:introduction}
\end{figure}

\minisection{LASAN}
The structure of Lightweight Authentication for Secure Automotive Networks (LASAN) is shown in \Cref{fig:introduction}.
A preliminary version of the protocols contained in the framework has been proposed in \cite{mundhenk2015a}.
These preliminary protocols were not secure and showed low performance.
The complete protocols are outlined in \Cref{sec:AuthAuth}.
It is optimized for the message distribution and architectures in the automotive context, specifically fixed networks and multicast messages.
The protocols have been extended based on the design considerations in this paper to guarantee real-time performance and security as shown in \Cref{sec:AuthAuth:extensions}.
LASAN further includes the protocols and procedures to be integrated with the automotive life cycle, making it a usable protocol.
These considerations are detailed in \Cref{sec:Integration}.
This enables necessary features required for the application of the protocol to real-world scenarios, such as the exchange of \glspl{ECU}.
LASAN is verifiable for security and can be evaluated regarding its real-time capabilities.

\subsection{Challenges and Opportunities}
\label{sec:Introduction:ChallengesAndOpportunities}
In automotive networks, the primary focus is on real-time capabilities to support control systems, which need to respond within a given short time, so predictability and reliability are the dominating factors.
At the same time, \glspl{ECU} have significantly less computational resources than modern computer systems.
Additionally, automotive networks have significantly lower data rates than their consumer counterparts.
Both of these limitations are motivated through the low cost nature of automotive components.
Maintaining predictability with \glspl{ECU} having low computational resources and low data rates while adding security is challenging.
Many existing security approaches assume greater computational capacity and higher bandwidth than is available in automotive networks and can thus not be applied to the automotive domain without adjustments.
To ensure security, as well as real-time behavior, we stage the required security actions into two phases: those that can be executed fast on \glspl{ECU} with low computational performance (symmetric operations) and those which require more computational effort (asymmetric operations).
We execute the slow, asymmetric operations when the vehicle is not in use, thus not interfering with real-time requirements.
The fast, symmetric operations in turn are executed when required and have minimal effect on the real-time performance of messages.
This allows the system to achieve the required real-time performance and maintain the desired level of security (see \Cref{sec:AuthAuth}).

Furthermore, it is important to note that automotive networks are defined at design time and typically do not change over the lifetime of a vehicle.
We utilize this fact by reducing flexibility in the security framework compared to similar frameworks on the Internet, based on the reasonable assumption that all devices run the same version of our proposed security framework.
Limiting flexibility and focusing on a static network allows us to reduce message sizes significantly, as no additional handshakes, version information, algorithm selection, etc. are required.
As these operations typically take a large amount of time, this reduction leads to a system with much lower latencies (see \Cref{sec:Evaluation}).

Lastly, messages in automotive networks are typically multicast or broadcast messages, reaching many or all receivers at the same time.
This behavior is not typical for consumer networks and thus, there is only limited support in existing security protocols.
We enhance existing protocols with the ability to authorize messages for multiple receivers, through the introduction of a root of trust in the network, the security module (see \Cref{sec:AuthAuth}).
This reduces the amount of required messages further, as only one handshake is required for all senders and receivers, instead of one handshake per sender/receiver pair.

A further challenge for security in automotive use cases is that external connectivity may be intermittent.
While cars are increasingly interconnected, this access is not available to all \glspl{ECU} in the network.
This is of course a security measure in itself, to shield \glspl{ECU} from potential attacks.
To overcome this challenge, we use asymmetric cryptography and certificates to allow components to make security decisions without requiring connectivity to external parties (see \Cref{sec:Integration}).

\subsection{Contributions}
\label{sec:Contributions}
Our detailed contributions are as follows:
\begin{enumerate}
	\item Our primary contribution is the development of Lightweight Authentication for Secure Automotive Networks (LASAN) (\Cref{sec:AuthAuth}).
	By addressing the challenges in \Cref{sec:Introduction:ChallengesAndOpportunities}, we create an efficient authentication and authorization framework, enabling secure real-time message transmissions.
	We show the full integration into the automotive life cycle (\Cref{sec:Integration}), thus, e.g., enabling secure exchange of \glspl{ECU}.
	To the best of our knowledge, LASAN is the only published protocol shown to be fully compatible with current automotive processes.
	\item While other frameworks have been aimed at security in the automotive domain, to the best of our knowledge, none of these has been formally shown to be secure.
	We thus formally establish the security properties of the protocols in LASAN using established protocol verification techniques.
	Specifically, we model and analyze the protocols using the Scyther tool \cite{cremers2008} (\Cref{sec:Verification}).
	\item We built a discrete event simulator (\cite{mundhenk2016SR}) for automotive networks and we use it to analyze the latencies and bandwidth requirements of LASAN with respect to the real-time constraints of this domain (\Cref{sec:Evaluation}).
	This simulator is further used to quantify the performance of LASAN in comparison to two existing authentication frameworks, \gls{TLS} and \gls{TESLA}.
\end{enumerate}
\section{Related Work}
\label{sec:RelatedWork}
%

In the following we give an overview of literature in the area of automotive security, including threats, countermeasures and integration.
We also explore adjacent domains to the automotive application.
Finally, we give a short overview of existing authentication frameworks in other domains.

\subsection{Automotive Threats}
Currently, most internal communication in vehicles is insecure.
Encryption is rare, and, if available, often uses similar keys across a series of vehicles and \glspl{ECU}.
Authentication is used primarily when reprogramming \glspl{ECU}, conventional data transmissions are not encrypted or authenticated.
The first extensive overviews of security in modern networked vehicles have been presented in \cite{koscher2010} and \cite{checkoway2011},
where it has been shown that \glspl{ECU} could be attacked and reprogrammed directly, by obtaining pre-programmed security keys from the car tuning community.
Most of these attacks have been performed with direct connections to the vehicle, but in \cite{checkoway2011} attacks via external interfaces such as the integrated telematics unit have also been reported.
More recently, in \cite{miller2013}, two vehicles have been attacked extensively and in \cite{miller2014}, multiple other vehicles have been analyzed.
The attacks by \cite{miller2013} have been executed over direct connections to vehicles, either to the \gls{OBD} port or the networks directly.
In \cite{miller2015}, these attacks have been extended and performed via a cellular connection.
Further examples of attacks, specifically for the \gls{CAN}, and countermeasures have been presented in \cite{hoppe2011}.
A case study investigating the likelihood of attacks on vehicles based on expert knowledge has been presented in \cite{benothmane2014}.

\subsection{Intrusion Detection, Network Analysis \& Verification}
In \cite{miller2014}, a vehicle \gls{IDS} has been proposed to detect anomalies in the vehicle network traffic by constantly comparing traffic to a baseline.
The approach in \cite{hoppe2008a} utilizes concepts like detection models known from the computer domain and port these to the automotive domain.
There, also the interaction with the driver is investigated, as responses from \glspl{IDS} in a safety-critical context can not always be automated.
Other approaches, such as \cite{muter2011} are based on the entropy of traffic, allowing a flexible detection of attacks without requiring predefined attack patterns.
Intrusion detection can be significantly more difficult when attackers do not perform their attacks in a straightforward fashion, but instead try to hide their intentions by obfuscating their attacks.

To validate the security of in-vehicle networks, \cite{sojka2014} propose a combined evaluation of safety and security requirements of an \gls{AUTOSAR} architecture.
In \cite{mundhenk2015}, probabilistic model-checking is used to evaluate the security of an automotive architecture, based on standard security assessment methods.

\subsection{Encryption and Hardware Support}
Initial efforts to introduce encryption into real-time vehicular networks have been presented in the literature \cite{sagstetter2013}.
To unify protection efforts, the \gls{HIS} has specified a cryptographic accelerator, called \gls{SHE} for use in vehicles~\cite{escherich2009}.
The \gls{SHE} is a cryptographic accelerator with cryptographic storage and is based on the EVITA \gls{HSM} light defined in the EVITA project \cite{seudie2009}.
Support for encryption and key storage can also be achieved with a \gls{TPM} \cite{iso-internationalorganizationforstandardization2009}.
Microcontrollers used in \glspl{ECU} are in many cases available with compatible hardware security modules with only minor price differences.
An alternative approach for integrating encryption into standard network controllers has been proposed in \cite{shanker2014} where this has been shown to have no impact on communication latency.
A fully featured implementation with symmetric cryptography and protocol obfuscation has been shown in \cite{shanker2015}.
However, these approaches rely on non-standard network controllers on FPGAs.

To limit the latency impact on the real-time communication, symmetric cryptography such as \gls{AES} \cite{nist2001} and the AES-based \gls{CMAC} \cite{dworkin2005} have been chosen in the \gls{SHE} and elsewhere in the literature, since these approaches are computationally simpler than asymmetric approaches.
These algorithms are significantly faster, especially with the hardware support of the \gls{SHE}.
Symmetric encryption requires that secret keys are known to all participants of a protected communication.
This opens a new attack vector, as keys are often pre-programmed into \glspl{ECU} and valid for the lifetime of the \gls{ECU}~\cite{koscher2010}.
In case a key becomes known, such as described in \cite{koscher2010}, attacks for the lifetime of vehicles or possibly on complete vehicle fleets are possible.
These scaling effects can be dangerous, as an attacker can leverage an attack on millions of vehicles.
With the more widespread use of vehicles in external networks, this recovery and publishing of secret keys will increase, as more groups can reap the benefits.
An approach to mitigate scaling effects through obfuscation of \gls{CAN} message IDs has been proposed in \cite{lms:2015}.



\subsection{Security Integration}
One approach to employing symmetric cryptography and limiting the overhead of additional security in legacy communication systems is to use Message Authentication Codes (MACs).
These are based on symmetric cryptography, allowing fast and efficient computation, especially on \glspl{ECU} with limited computational power.
In \cite{lin2013}, MACs are introduced to \gls{CAN} and safety and security are considered in an integrated functional model mapped to a \gls{CAN} architecture.
In \cite{han2014}, an approach to introduce MACs into FlexRay is presented.
There, the \gls{TESLA} protocol is employed for time-delayed release of keys in the time-triggered segment of FlexRay.
While \gls{TESLA} supports sender authentication with symmetric mechanisms, it does not authenticate communication partners or authorize communication streams.
In \cite{zalman2014}, a \gls{CMAC} is utilized and integrated with the \gls{CRC} of the underlying communication system, reducing the overhead and achieving the required sender authentication and message integrity checks.
By including the security considerations in the design phase of systems, \cite{jiang2012} achieve time efficiency with low hardware overhead.
To be able to ensure real-time behavior, all of these approaches employ symmetric cryptography, requiring a pre-shared key.

Others, such as \cite{herber2014} propose the use of Virtual CANs (VCANs), similar to virtual networks in the consumer and corporate domain, to separate network traffic.

\subsection{Other domains}
While the abovementioned approaches focus on in-vehicle networks, the adjacent domain of car-to-x (car-2-x, c2x) technologies has received more extensive coverage in the literature.
Noteworthy here is the \gls{SeVeCom} project \cite{kargl2009}.
This outlined a first attacker model and proposed security mechanisms for c2x networks.
\gls{SeVeCom} relies on multiple levels (long-term, short-term) of public-key mechanisms.
This approach has been used in other projects and multiple consortia are now working on c2x technologies.
However, participants need sufficient computational resources for asymmetric cryptography to implement these systems.

Significant progress in securing communication has been achieved in other areas where embedded systems are used to transmit data.
The smart grid is a prime example of progress in this domain.
Multiple approaches have been described in literature to achieve this.
In \cite{paverd2014a} a system-level approach to secure smart meters has been proposed, based on the hardware support presented in \cite{paverd2012}.
In \cite{sikora2013a} an approach to secure communication for the smart grid has been proposed.
There, \gls{TLS} has been used to secure the communication of the backbone.
However, in comparison to smart metering applications, the real-time constraints in the automotive domain are more stringent.

\subsection{Authentication Frameworks}
\label{sec:AuthenticationFrameworks}
To reduce the risk from pre-shared keys, authentication frameworks have been developed in other domains, such as corporate networks and the Internet, allowing the exchange of keys without prior interaction of the communication participants.

\minisection{Kerberos and TLS}
Examples include Kerberos~\cite{neuman2005} and the widely used \gls{SSL}/\gls{TLS} framework~\cite{dierks2008}.
Kerberos can be extended to a two-phase system to initialize the system with public-key mechanisms~\cite{zhu2006}.
These mechanisms have been designed for use in computer networks and have significant overheads, preventing their efficient use in real-time systems.
They allow the exchange of keys without prior knowledge of the communication participants by adding an element of trust in the network.
This can be a trusted server on the Internet or a specially secured \gls{ECU} in vehicle networks.
However, as we will show, the real-time performance of such mechanisms, as well as their fundamentally different orientation towards Internet communications makes these frameworks unsuitable for the automotive domain (see \Cref{sec:Evaluation}).

\minisection{TESLA}
The \gls{TESLA} system has been designed specifically for low-performance communication systems \cite{perrig2005}.
In \gls{TESLA}, the sender generates a new symmetric key and computes the MACs for one or more messages.
After the messages have been received, the sender broadcasts the key on the bus, allowing every recipient to authenticate the sender of the previous message.
The communication overhead of \gls{TESLA} is minimal, especially since the keys are sent alongside the data of the next message.
However, the unavoidable time delay between receiving and authenticating a message limits TESLA's applicability in real-time systems.
Groza et al. \cite{Groza2012} argue that this delay is too large for intra-vehicle communication scenarios.
Further, \gls{TESLA} only supports limited receiver authentication in communication systems without sender identifiers, and does not provide stream authorization or encryption.


\minisection{LiBra-CAN}
For the automotive domain, LiBra-CAN is presented in \cite{Groza2012}.
It authenticates senders at the receiving \glspl{ECU} via \glspl{M-MAC}.
Keys are assigned to groups of \glspl{ECU}.
LiBra-CAN requires pre-shared keys and does not concern itself with key exchanges.

\minisection{CANAuth}
In \cite{van2011canauth}, CANAuth, a very lightweight authentication mechanism, is proposed.
CANAuth allows broadcast authentication and keys are assigned for message groups.
Similar to LiBra-CAN, CANAuth requires pre-shared keys, before authentication can be performed for message groups.

\minisection{VeCure}
More recently, VeCure has been proposed in \cite{7030108}.
VeCure splits \glspl{ECU} into classes based on trust and assigns keys to these classes.
However, VeCure also relies on pre-programmed keys and proposes to program these at the initial setup of the vehicle.

Most frameworks in literature rely on pre-programmed keys.
However, such approaches are not realistic, sufficient or effective.
The secure generation and the programming of keys is not discussed and not a trivial problem.
Additionally, due to the long lifetime of vehicles, keys must be renewed regularly, to avoid attacks.
Further, the exchange of \glspl{ECU} in workshop situations is difficult or even impossible.
Thus, the integration aspects are not considered here, making the protocols unsuitable for real-world applications.


\section{Authentication \& Authorization}
\label{sec:AuthAuth}
In this section, we introduce the \gls{ECU} authentication (\Cref{sec:Design_ECUAuthentication}) and stream authorization (\Cref{sec:Design_StreamAuthorization}) approaches in detail.
The preliminary concept of these two protocols was proposed in \cite{mundhenk2015a}.
Through thorough protocol analysis and prototype implementation, a number of messages have been reformulated for increased security and reduced latency (see \Cref{sec:AuthAuth:extensions}).

\subsection{Terminology}
\label{sec:AuthAuth:terminology}
The following terminology is used to describe the framework.
We denote an \gls{ECU} as $e$, sending a set of streams $S_e$ to a set of receiving \glspl{ECU} $R_e$.
In turn, a stream $s \in S_e$ is identified by the sending \gls{ECU} $e$, a set of receiving \glspl{ECU} $R_s$ and a set of message instances $M_s$: $s=(e,R_s,M_s)$.
A message instance $m \in M_s$ is assigned to a stream $s$ and carries a payload $o$.
The security module of the system is denoted as $y$.
A timestamp on device $d$ is denoted as $\omega_d$ and a random number as $\rho$.
A nonce $n$ is a unique random number $\rho$ within the accuracy of one timestamp.
Together, a timestamp and nonce shall be unique over the lifetime of the vehicle, while minimizing the storage required for nonces.
A key is denoted as $k$ and identified by its value $v$ and its length $l$.
We further define a hashing function for message $m$ on device $d$ as $h_d(m)$.
A parameter to determine if the system is running is defined as $\theta \in \{0,1\}$ and a function $\alpha(e,s) \in \{0,1\}$ determines if an \gls{ECU} $e$ has access to the stream $s$ (1) or not (0).
The time required to execute a function $x$ is defined as $\tau_x$.
An action $z$ is triggered, if a condition $i$ is fulfilled: $i \rightarrow z$.

\subsection{ECU Authentication}
\label{sec:Design_ECUAuthentication}
\begin{figure}
	\centering
	\includegraphics[width=\hsize]{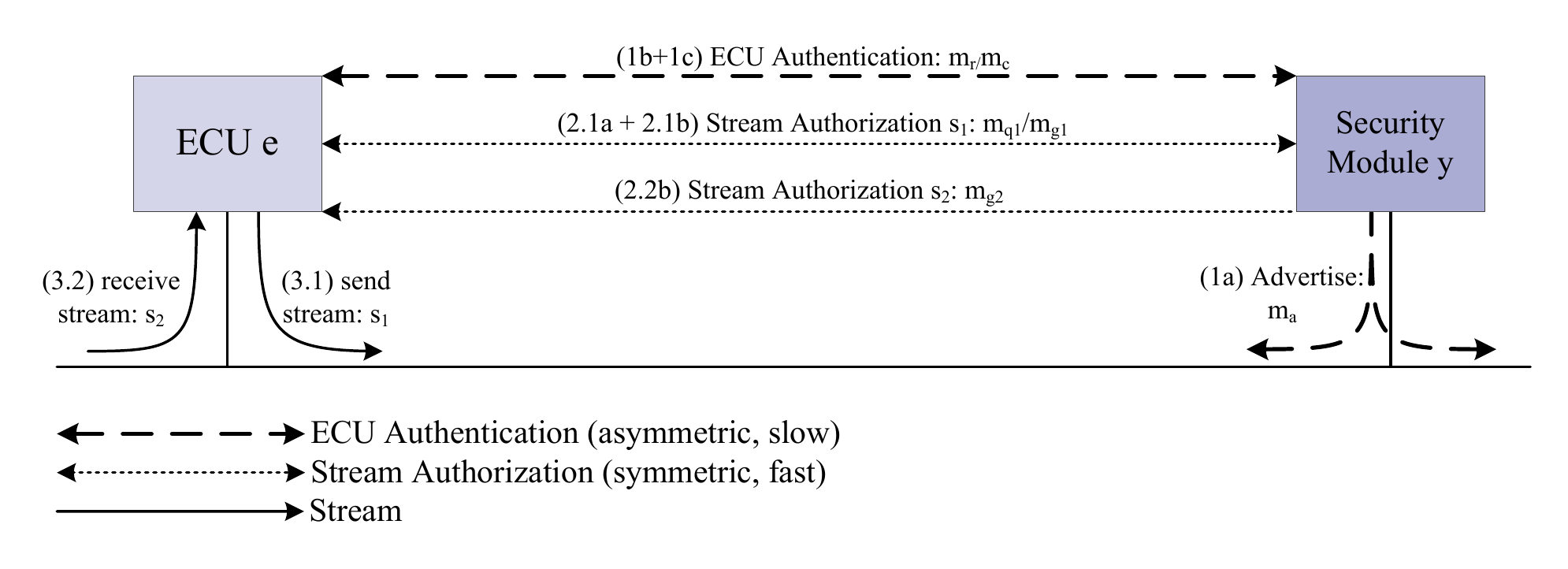}
	\caption{In our framework, every ECU is authenticated against the security module (1a-1c).
	Subsequently, the ECU can request the keys for a message stream (2.1a-2.1b) and start transmitting (3.1).
	If the ECU is to start receiving a message stream, it is notified by the security module with the message stream key (2.2b) before it receives the stream (3.2).}
	\label{fig:overview}
\end{figure}

The key exchange used for \gls{ECU} keys is based on the \gls{PKINIT} protocol for Kerberos.
\gls{PKINIT} has been chosen for its conciseness.
This reduces the required time and bandwidth for the authentication.
The \gls{ECU} authentication is illustrated in \Cref{fig:overview} and described in the following.
This mechanism facilitates the initial key exchange between the \gls{ECU} and the security module, and is based on asymmetric cryptography.
Secure asymmetric cryptography is generally a computational expensive task.
This holds especially for software implementations without hardware support, as will be the case on most \glspl{ECU}.
Thus, this \gls{ECU} authentication needs to be executed when the vehicle is not in use as not to interfere with the real-time capabilities of the system.

\minisection{Authentication Mechanism}
To describe \gls{ECU} authentication, we define a set of keys for device $d$, consisting of a public key $k_{d,\mathrm{pub}}$ and a private key $k_{d,\mathrm{priv}}$.
We further define an asymmetric encryption function $\epsilon^a$ for device $d$, that translates clear text $t$ into cipher text $c$ with key $k$, such that $c = \epsilon^a_d(t,k)$.
Accordingly, we define an asymmetric decryption function $\delta^a$ for device $d$, that translates cipher text $c$ into clear text $t$ with key $k$, such that $t = \delta^a_d(c,k)$.
For authentication, we further require the certificate $f_d$ for device $d$ and a function $\phi_{d_{1}}(f_{d_{2}}) = \{0,1\}$ to verify the certificate $f_{d_{2}}$ on device ${d_{1}}$.

As shown in \Cref{fig:overview}, three steps are required to authenticate an \gls{ECU}.
First, the security module needs to make itself known and authenticate to all \glspl{ECU}.
Second, each \gls{ECU} authenticates to the security module and requests a symmetric key.
Third, the security module permits the requesting \gls{ECU} to access the bus by sending a confirmation message to this \gls{ECU}.
These steps are described in the following:

\begin{enumerate}[(a)]
	\item
Advertisement by the security module (\Cref{fig:overview}~(1a)):
The security module $y$ advertises its certificate to every \gls{ECU} on the bus at startup with message $m_a$.
This certificate is required to be signed by the appropriate \gls{CA} (see Section~\ref{sec:CRL}).
\begin{align}
\label{equ:ad}
	\theta \rightarrow m_a \ \mathrm{with} \ m_a \in M_{s}^{a}, \ s = (y, \{e\},M_{s}^{a}), \ o = f_{y}
\end{align}
The authentication begins with the security module $y$ presenting its security certificate $f_y$ to \gls{ECU} $e$.
This certificate includes the security module's public key, and is broadcast on the network unencrypted.
Each \gls{ECU} has a list of trusted \glspl{CA}, which it uses to verify the signature on the security module's certificate.
This list of trusted \glspl{CA} can be updated by the manufacturer using the remote software update procedure (see Section~\ref{sec:Scen_OTA}).
	\item
Registration of the \gls{ECU} (\Cref{fig:overview}~(1b)):
Once the security module has been successfully authenticated ($\phi_{e}(f_y) = 1$), the \gls{ECU} transmits its registration message $m_r$, including its own certificate $f_e$ and the newly generated \gls{ECU} key $k_{e,\mathrm{sym}}$.
The key, a timestamp $\omega_{e}$ and a nonce $n$ are encrypted with the public key of the security module $k_{y, \mathrm{pub}}$ and can thus only be decrypted by the security module.
The certificate $f_e$ contains the ECU's identity and public key, as well as any manufacturer-defined properties of the ECU (e.g. type of ECU, current software version):
\begin{align}
\label{equ:register}
	\phi_{e}(f_y) &\rightarrow m_r\\
	\mathrm{with} \ m_r &\in M_{s}^{r},\ s = (e, \{y\}, M_{s}^{r}),\nonumber\\
	o &= (\epsilon^{a}_{e}((e, y, k_{e, \mathrm{sym}}, n, \omega_{e}), k_{y, \mathrm{pub}}), \epsilon^{a}_{e}(h(e, y, k_{e, \mathrm{sym}}, n, \omega_{e}),k_{e, \mathrm{priv}}), f_{e})\nonumber
\end{align}

	\item
Confirmation of the security module (\Cref{fig:overview}~(1c)):
Upon receiving the registration message, the security module authenticates the ECU by decrypting the received hash from $m_{r}$ and comparing it to the hash of the received elements.
The elements have been received separately in $m_{r}$ and need to be decrypted with the private key of the security module $k_{y,\mathrm{priv}}$.
Additionally, the certificate of the sender is verified and, in the case of successful verification ($\phi_{y}(f_e) = 1$), the security module saves the \gls{ECU} key, the \gls{ECU} identity, and any manufacturer-defined information.
The security module then sends a confirmation message $m_c$ to the \gls{ECU}.
This message contains the \gls{ECU} identifier encrypted with the newly exchanged symmetric \gls{ECU} key:
\begin{align}
\label{equ:confirm}
	\delta^{a}_{y}(\epsilon^{a}_{e}(h(y, &k_{e, \mathrm{sym}}, n, \omega_{e}),k_{e, \mathrm{priv}}), k_{e, \mathrm{pub}})=h(y, k_{e, \mathrm{sym}}, n, \omega_{e})) \wedge \phi_{y}(f_e) \rightarrow m_c\\
	\mathrm{with} \ m_c &\in M_{s}^{c}, \ s = (y, \{e\}, M_{s}^{c}), \ o = (e, y, \epsilon^{s}_{y}((e, n, \omega_{y}), k_{e, \mathrm{sym}}))\nonumber
\end{align}
\end{enumerate}
In addition to the certificate and signature verification steps described above, the validity of all timestamps is checked for every message.

In order to transmit messages via an encrypted channel, the \gls{ECU} must request a stream key from the security module.
This is handled via the stream authorization protocol, as described in the following.

\subsection{Stream Authorization}
\label{sec:Design_StreamAuthorization}

After an \gls{ECU} has authenticated itself to the security module and established a symmetric \gls{ECU} key, the \glspl{ECU} can request initiations of message streams.
All message streams and stream keys in the system are managed by the security module.
Each message stream uses a unique (within the network) stream key, which the sending and receiving \glspl{ECU} request from the security module.
By centralizing ECU authentication and stream authorization at the security module, our framework eliminates the need for the sending and receiving \glspl{ECU} to authenticate each other.
The stream setup can be performed when a message stream is required, or, if the real-time requirements are more stringent than the stream setup time, considerably in advance (e.g. at vehicle start-up).

\minisection{Authorization Mechanism}
For stream authorization, we define a symmetric key $k_{\mathrm{sym}}$ and two functions $\epsilon^s$ and $\delta^s$ for encryption and decryption, such that $c = \epsilon_{d}^{s}(t,k_{\mathrm{sym}})$ and $t = \delta_{d}^{s}(c,k_{\mathrm{sym}})$ for plaintext $t$ and ciphertext $c$ on device $d$.
\ \\
The stream can only be established after the \gls{ECU} receives the confirmation message $m_c$ from the \gls{ECU} authentication protocol (see \Cref{fig:overview} (1c)).
The stream is usually established when a message $m$ is requested to be sent.
The sending \gls{ECU} $e$ requests a key from the security module $y$, allowing access to the stream.
The request message $m_q$ contains the identifier of the requested stream $s_{i}$, the identifier of the requesting \gls{ECU} $e$, as well as a timestamp $\omega_e$ and a nonce $n$ to protect against malicious retransmissions.
The content of the message $m_q$ is encrypted with the \gls{ECU} key $k_{e, \mathrm{sym}}$:
	\begin{align}
	\label{equ:stream}
		m_c \wedge m &\rightarrow m_q\\
		\mathrm{with} \ m_q &\in M_{s}^{q}, \ s = (e, \{y\}, M_{s}^{q}),\nonumber\\
		o &= (e, \epsilon^{s}_{e}((e, s_{i}, n, \omega_e), k_{e, \mathrm{sym}}))\nonumber
	\end{align}
If the received request message $m_q$ can be successfully decrypted with the \gls{ECU} key $k_{e, \mathrm{sym}}$, it must have originated from the correct ECU.
The security module can then make an access control decision based on the stored information about that particular \gls{ECU}.
The security module can support arbitrarily-complex access control policies defined by the manufacturer.
One such policy could be for the security module to maintain an access control list (ACL) based on the \gls{ECU} identities.
If the security module determines that the requesting \gls{ECU} $e$ is allowed access to the stream $s_{i}$ ($\alpha(e, s_{i}) = 1$), the security module $y$ assigns a new stream key $k_{s_{i}, \mathrm{sym}}$.
This key is first sent to all receiving \glspl{ECU} in the stream $\tilde{e} \in R_{s_{i}}$, and then to the requesting sending \gls{ECU} $e$.
In this way, the protocol ensures that all receivers have access to the data as soon as it is sent:
	\begin{align}
	\label{equ:grant}
		m_q \wedge (\delta_{y}^{s}(m_q, k_{e, \mathrm{sym}}) &= (e, s_{i}, \omega_e, n))\\
		\wedge (\omega_e \leq \omega_y) \wedge \alpha(e, s_{i}) &\rightarrow \forall \tilde{e} \in \{R_{s_{i}}, e\}: m_g\nonumber\\
		\mathrm{with} \ m_g &\in M_{s}^{g}, \ s = (y, \tilde{e}, M_{s}^{g}),\nonumber\\
		o &= \epsilon^{s}_{y}((\tilde{e}, s_{i}, k_{s1, \mathrm{sym}}, n_i, \omega_y), k_{\tilde{e}, \mathrm{sym}}),\nonumber\\
		n_i = \rho : \rho \notin &\bigcup_{l} \rho_{l}, l=\{0..i-1\}\nonumber
	\end{align}
In theory, it would be possible to allow access to any \gls{ECU} of the correct type (e.g. motor control) that presents a valid certificate.
However, it must be assumed that some \gls{ECU} keys may have been compromised by the adversary.
Since it is not always possible to check the validity of a given certificate, using an ACL based on ECU type could lead to widespread attacks if the adversary extracts a key from a popular type of ECU (e.g. infotainment).
The use of individual ECU identities in the ACL limits the scope of such an attack to a single vehicle (usually the adversary's own vehicle).

\subsection{Extensions}
\label{sec:AuthAuth:extensions}
The above descriptions include significant revisions to the concept as presented in \cite{mundhenk2015a} as a result of a detailed evaluation using the protocol analysis tool (as described in \Cref{sec:Verification}).
These enhance security significantly and also address inefficiencies when evaluating the real-time performance as shown in \Cref{sec:Evaluation}.
LASAN addresses these identified security issues and optimizes the protocols.

Specifically, we found that in \cite{mundhenk2015a}, the \glspl{ECU} are not sufficiently secured against impersonation by the adversary.
The adversary could impersonate the \gls{ECU} and, through the exchanged \gls{ECU} key, gain authorization for potentially malicious messages.
This issue has been rectified through the introduction of an additional encrypted hash in the registration message $m_r$ (see \Cref{equ:register}): $\epsilon^{a}_{e}(h(e, y, k_{e, \mathrm{sym}}, n, \omega_{e})$.
This does entail increased latency, since this hash is encrypted with the private key of the requesting \gls{ECU}, and hence, the generation of this message takes significantly longer (see \Cref{tab:RSA}).
However, the use of the private key is unavoidable as it is the only way to securely identify and authenticate the \gls{ECU}.

When implementing the concept and evaluating its real-time capabilities, as shown in \Cref{sec:Evaluation}, we identified inefficiencies in the \gls{ECU} confirmation message $m_c$, as well as in the stream request message $m_q$.
Since these messages are encrypted, it is not possible for a device to determine whether or not it is the intended communication partner without first attempting to decrypt the message.
In the case of the confirmation message $m_c$, every \gls{ECU} first must decrypt the message with its private key, before being able to judge if it is the intended receiver.
Similarly, for the request message $m_q$, the security module must test all available ECU keys to decrypt the message.
To avoid these inefficiencies, we introduced an additional, unencrypted \gls{ECU} identifier $e$ into each of the messages.
Although this allows the adversary to determine the message’s recipient, this is not generally considered to be security-sensitive information in this domain.
Unless the adversary can subvert the cryptographic algorithms themselves (which is assumed to be beyond the adversary’s capabilities), the desired security properties are maintained.

\section{Integration}
\label{sec:Integration}
As for every security measure, the integration is an important factor to guarantee the security.
In the following, we describe the handling of certificate validations, as well as scenarios outside the typical application of LASAN, such as the setup of the vehicle.
These aspects are an essential part of every authentication framework to ensure full vehicle functionality.
However, in literature, these aspects are often not covered, also because their implementation is not trivial, often leading to suboptimal implementations and security flaws.

\subsection{Certificate Validation}
\label{sec:CRL}
The security of digital certificates depends on the security of the corresponding private keys.
Any entity that knows the relevant private key can claim the identity and attributes described in the certificate.
Therefore, if a private key is stolen or purposely distributed, the corresponding certificate can no longer be trusted.
For example, if the private key of an ECU is stolen, a malicious device can impersonate that specific ECU to the rest of the vehicle network.
Although ECU manufacturers can take measures to protect the private keys, it must be assumed that some private keys might be extracted from ECUs.
In particular, an adversary with the right tools and physical access to an ECU (e.g. an after-market component) would likely be able to extract the private key.
In light of this, our system is designed to minimize the impact of a compromised private key, as far as possible.

The same challenges regarding compromised private keys also arise on the Internet and in other traditional computer networks that make use of digital certificates.
In those contexts, approaches such as \glspl{CRL} \cite{cooper2008} and the \gls{OCSP} \cite{santesson2013} are used to address this issue.
However, neither of these approaches is suitable for use in vehicular networks, due to the unique characteristics and constraints of this domain.
CRLs are essentially lists of certificates that have been revoked (e.g., because of a known compromise of the private key).
In order to be effective, CRLs must be frequently updated and checked whenever a certificate is verified.
Although many modern vehicles are likely to have connectivity to external networks, it cannot be assumed that this connectivity will be available at all times (e.g., the vehicle may lose connectivity if traveling in a remote area).
Furthermore, CRLs on the Internet usually contain a large number of revoked certificates, making it burdensome for the vehicle to frequently download these relatively large data sources and to store and process these lists when checking certificates.
To avoid the need to download large CRLs, OCSP allows a verifier to query the status of an individual certificate, to ascertain whether or not it has been revoked.
However, this requires always-on connectivity in order to check individual certificates, which again may not always be available.
Furthermore, both of these techniques present the adversary with new vectors to mount denial of service attacks against the vehicle network.
For example, if the adversary can change the OCSP response or cause the vehicle to download a modified CRL, the vehicle may refuse to accept certificates from legitimate ECUs.
Although this may not be a serious issue on the Internet, this type of attack would have a much greater impact in the vehicle network context.
A potential attack could be the disabling of the brake controller and thus the brakes at high-speeds.

Instead of using CRLs or OCSP, we have designed a new protocol specifically for use in vehicle network contexts.
The main ideas behind our certificate validation protocol are as follows:

\begin{itemize}
\item The vehicle manufacturer should not be required to provide an online service (e.g. OCSP) since this may incur high operating costs, especially because it would have to be available for the full life of all vehicles on the road.
\item The protocol does not attempt to prevent owners from intentionally installing modified/rogue ECUs in their own vehicles. However, in the event of an investigation, this should be detectable in order to ascertain liability in case of vehicle malfunction.
\item The protocol only takes place when an ECU is added/exchanged, since this is when the system is most vulnerable.
\item The validation step relies on a human-in-the-loop, who could either be the vehicle owner or a trusted representative (e.g. a vehicle workshop).
\item The protocol must not restrict the choice of workshops/service centres at which the vehicle may be serviced, since this may be considered anti-competitive behaviour.
\end{itemize}

The full details of our certificate validation protocol are presented in the system life-cycle scenarios in the next section.

\subsection{System Life-Cycle Scenarios}
\label{sec:Scenarios}
Throughout its lifetime, every vehicle experiences a certain set of scenarios, which are divergent from the standard driving behavior.
It is essential for any authentication protocol targeted to the automotive domain to address these use cases.
Thus, to ensure full operational capabilities, these situations must be handled in LASAN.
In this section, we describe these scenarios, as well as the support by LASAN.

\subsection{System Setup}
\label{sec:Scen_SystemSetup}
\begin{figure}
	\centering
	\includegraphics[width=\hsize]{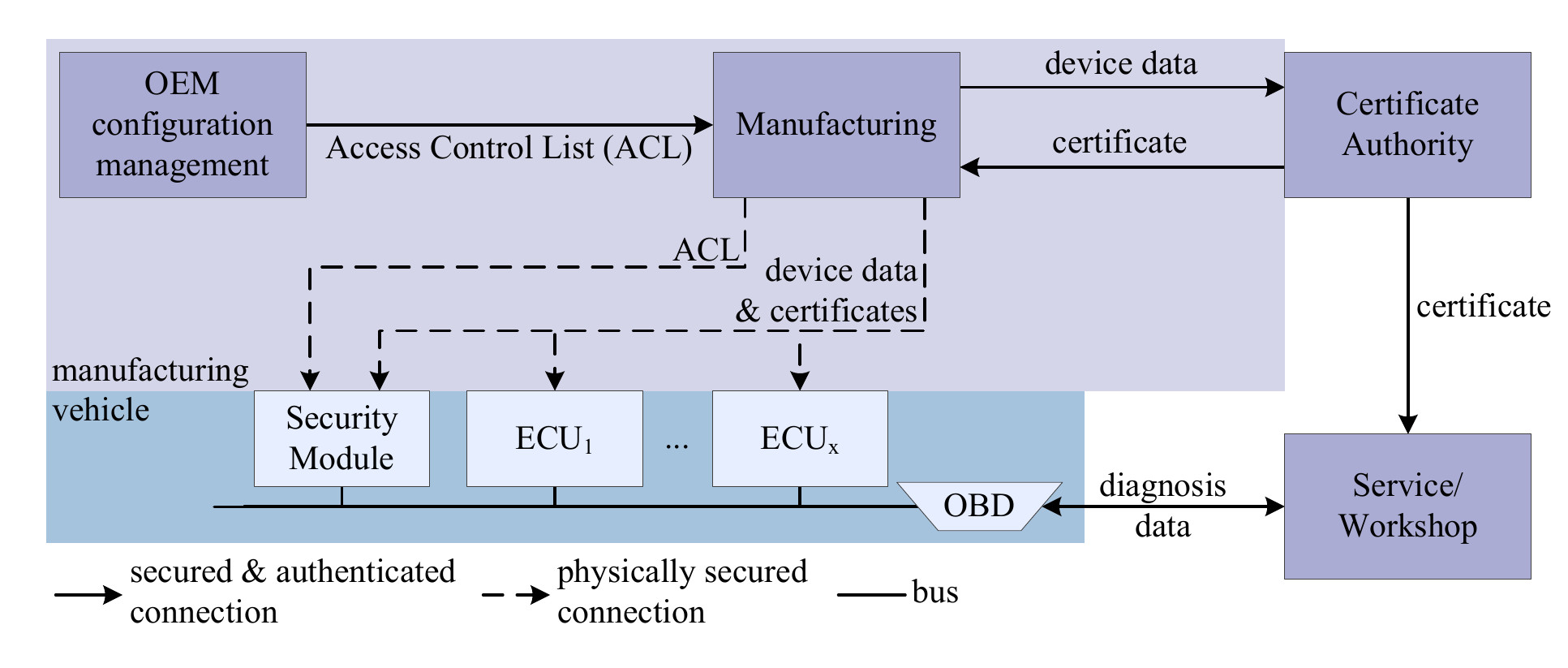}
	\caption{Overview of system setup and workshop situations. At manufacturing, every device is programmed an ID and a certificate. Security modules are additionally supplied the Access Control List (ACL) for the vehicle. A workshop connects to the vehicle via a secured connection to the On-Board Diagnosis (OBD) port. All participants are certified by the central Certificate Authority (CA) of the Original Equipment Manufacturer (OEM).}
	\label{fig:setup}
\end{figure}

The first non-standard situation the vehicle experiences is the manufacturing of its components and the assembly of the vehicle in the factory.
This process is shown in \Cref{fig:setup}.
At some point in this process, the \glspl{ECU} and security module are bare and do not have any security certificates or keys.
The security of all devices and connections in the following process is essential to guarantee the security of subsequent states.

Where possible, the asymmetric key pairs are generated by the relevant devices (i.e. the security modules and ECUs).
However, where this is not possible, these keys are generated externally and securely provisioned onto the devices.
To facilitate the certificate validation protocol, each device is assigned a globally-unique human-readable ID (e.g. a short alpha-numeric code), which is printed on the physical device.
Each device is issued a digital certificate that contains its device ID, its public key, and any device-specific information (e.g. type of device and operating parameters).
This certificate is signed by the respective Certificate Authority (\gls{CA}), which could be the device manufacturer.
This certificate binds the device ID to its public key, and thus only a device in possession of the corresponding private key can assert that identity.
During this process, the programming connection between computer and device must be secured both digitally and physically.
This connection can be placed in the manufacturing process of the device, ideally at the first start of the device (\gls{ECU} or security module).

As explained in the previous section, the security module holds an \gls{ACL} containing all permissible message streams.
The IDs of the authorized sending and receiving ECUs for each stream are programmed into this ACL.
With this list, the security module can verify if all \glspl{ECU} are correctly built into the vehicle and which \glspl{ECU} have access to which streams.
The list can be automatically generated by the configuration management of the \gls{OEM} and must be signed with the corresponding certificate.
Note that an \gls{ACL} is a feasible method here, as the distribution of tasks and messages is known at design time.
In case this changes via a functionality update, this may also contain an update to the \gls{ACL}.

\subsection{Vehicle Service}
\label{sec:Scen_VehicleService}

Although a vehicle network may not change frequently, there will always be a need to replace certain components in the system.
For example, this might happen when a damaged \gls{ECU} needs to be replaced.
To the best of our knowledge, LASAN is the only framework supporting such exchanges.
In the following, we describe the protocol to exchange \glspl{ECU} but omit message details for conciseness.

The two objectives of this protocol are (1) to determine whether or not a particular ECU is authentic and, (2) to securely add its ID to the ACL held by the security module.

The first objective is achieved using the ID printed on the physical device and makes use of manual validation as follows:
When a new ECU is ready to be added, the vehicle owner or workshop attempts to determine whether the same ECU ID appears in any other vehicle.
If it does, this means that there is more than one ECU with access to that particular private key, indicating that the key has been compromised.
To facilitate the process of checking for duplicate IDs, vehicle owners and workshops should be encouraged to publish the lists of ECU IDs they have installed.
These public lists can then be consulted to check for duplicate IDs.
This process is similar to the concept of Certificate Transparency, recently developed and deployed by Google~\cite{rfc6962}.

Once satisfied that the ECU is authentic, the vehicle's owner puts the vehicle into a special state in which new ECU IDs can be added to the security module's ACL.
Various different approaches could be used to authenticate the owner and put the vehicle into this state, including a physical switch, a special vehicle key, an owner password, or a combination of these factors.
Once in this state, the owner or workshop can add the ECU and verify that the digital ID matches the ID printed on the physical device.

It is a deliberate design decision that this procedure is not fully automated, and we argue that this level of manual intervention is both feasible and necessary, given that vehicle service is not likely to be a frequent occurrence.
Since the vehicle owner authorizes this process, there is no need to contact the manufacturer.
This also means that the owner may permit any third party to perform the vehicle service (e.g. the vehicle is not restricted to manufacturer-approved service stations).
The security module keeps a record of the ECU IDs that have been added in this way, and this record can be audited if need be (e.g. as part of an investigation into vehicle malfunction).

\subsection{Firmware updates}
\label{sec:Scen_OTA}
In connected vehicles, the possibility of \acrfull{OTA} updates is becoming increasingly important.
For unconnected vehicles, firmware updates are usually performed at workshops.
By being able to update software at any point in time, the number of visits to repair shops, as well as the downtime of the vehicle are reduced significantly.
However, firmware updates, especially \gls{OTA} updates, can compromise security, if not handled properly.
\gls{OTA} updates are available in some vehicles on the market today.
In these cases, signed firmware updates are used and are verified by the receiving \gls{ECU}, if possible.
If the receiving \gls{ECU} does not have the computational power to verify the certificate, an additional element, such as the cellular gateway can perform this task.
In this case, the programming of the \gls{ECU} is performed over the internal networks without security, based on trust.
The only security measure against illegitimate firmware updates from the \gls{CAN} bus is usually a simple password that can often be recovered from the tuner community \cite{koscher2010}.

In LASAN, we can replace this simple password with our authentication scheme and ensure that the firmware update is authenticated all the way to the \gls{ECU}.
All firmware updates are run via the security module, which in turn verifies the identity of the sender and the validity of the firmware signature for the \gls{ECU}.
The firmware is signed with the private key of the sending instance, which is certified by an appropriate \gls{CA} (see \Cref{sec:CRL}).
After successful validation, the security module triggers the programming of the \gls{ECU}.
To ensure a maximum of security, the \gls{ECU} authentication (see \Cref{sec:Design_ECUAuthentication}), including the re-validation of the security module certificate, is started as soon as a reprogramming command is received and before a new firmware is accepted.
After successful authentication of both the security module and the \gls{ECU}, the update is installed on the \gls{ECU}.

In this Section we presented an essential part of LASAN, the integration with the processes in the automotive domain.
Although often omitted in literature, the protocols and considerations described are necessary requirements for any authentication framework design, as they represent the use cases which must be supported by the authentication framework.
\section{Verification}
\label{sec:Verification}
In this Section, we focus on the modeling and security verification of the protocols contained in the framework with the protocol verification tool Scyther \cite{cremers2008}.
We provide a brief introduction to Scyther and define our adversary model.
As it is not possible to verify the overall framework, we separately verify the protocols contained in LASAN.
For this, we do require a set of assumptions, which we will detail here.
We then present the formal models of the protocols and the results of the verification.
As conceptual security is paramount in authentication frameworks, the models developed in the following are available online for public scrutiny \cite{ResultsRepo}.

\minisection{Scyther Tool}
The Scyther tool, developed by Cremers \cite{cremers2008,cremers2008a}, performs automatic symbolic analysis of security protocols in terms of their confidentiality (secrecy) and authentication properties.
In Scyther, protocols are modeled as sets of \emph{role scripts} which are executed by \emph{entities}.
Each execution of a role is called a session and multiple concurrent sessions are permitted for any role.
The desired security properties for a protocol are modeled as \emph{claims} made by specific roles.
If a claim is falsified, Scyther offers a counterexample showing the sequence of events leading to that point, which will usually represent a potential attack against the protocol.
For certain claims, Scyther can perform \emph{unbounded verification} which, if successful, indicates that the claim will not be falsified irrespective of how many times the protocol is run \cite{cremers2008a}.
Internally, Scyther uses the concept of \emph{patterns} to reason about infinite sets of traces in a protocol \cite{cremers2008a}.
A pattern forms a labeled directed acyclic graph that, under certain conditions, could be a realization of the protocol under analysis.
Patterns can be \emph{refined} according to a set of rules and well-typed substitutions.
The verification algorithm applies a pattern that violates a specific claim and attempts to refine this into a realizable pattern that then represents a counterexample to the claim.

\minisection{Adversary Model}
In line with the majority of security protocol analyses, we assume a Dolev-Yao adversary model \cite{dolev1981}, in which the adversary has full control over all communication and is thus able to intercept, modify, replay or block any messages, as well as inject new constructed messages.
This is appropriate for our application domain since the adversary could, for example, be a malicious device connected to the communication bus which, in the worst case, would be the central gateway, able to exert this degree of control over all messages on the bus.
In the model, the adversary can also perform any role, which in practice means that the adversary could pretend to be any device on the bus.
However, the Dolev-Yao adversary is assumed to be computationally bounded and thus unable to break cryptographic primitives (encryption, digital signatures, hash functions etc.) in a reasonable amount of time.
This is a realistic assumption, given that these cryptographic primitives are widely used in other domains (including Internet communication) and are not known to be broken.
Finally, it is always possible that a real-world adversary might be able to compromise one of the legitimate devices (e.g., through malicious software or runtime exploits).
However, since this is almost entirely dependent on the actual hardware and software implementations, this class of attacks is beyond the scope of our analysis and we assume that an authenticated device is secure.


\minisection{ECU Authentication}
The model of the ECU authentication protocol is described in \Cref{sec:Design_ECUAuthentication}.
While the design of the framework includes the distribution of certificates in messages $m_a$ (\Cref{equ:ad}) and $m_r$ (\Cref{equ:register}), these certificates are omitted in the Scyther model.
The verification of certificates depends on external knowledge, such as locally available root certificates or connections to external verification agencies, such as the certificate \gls{CA}.
This exceeds the complexity that Scyther can verify and certificates are thus omitted in our analysis.
As methods for initial certificate distribution and verification have been used successfully and securely for many years on the Internet, these can be applied here as well and the omission is reasonable.

To verify security, we define a set of claims which, if they hold, ensure the security of the protocol.
This step is crucial, as insufficient claims would lead to security flaws being ignored by Scyther.
Each role makes the following two security claims after the protocol:

\begin{enumerate}

\item Secrecy of symmetric \gls{ECU} key: If neither the ECU's nor the security module's private keys have been compromised, then the newly established shared key cannot be known by the adversary. This is due to the fact that the only way to recover the shared key is by decrypting it with the private key of either component.

\item Non-injective synchronization: On completion of a protocol run, all entities agree on the roles that have been taken and the data items that have been exchanged.
This synchronization is \emph{non-injective} in the sense that replays of complete protocol runs are excluded since these would be detected using the nonces and timestamps in the messages.

\end{enumerate}

Scyther confirms that both of these claims are successfully verified in the unbounded sense from both the perspective of the ECU and the security module.
This means that under the defined Dolev-Yao adversary model, this protocol ensures the confidentiality of the newly-generated shared key and achieves mutual authentication.

As mentioned in \Cref{sec:AuthAuth:extensions}, the signed hash term in message $m_r$ is critical for achieving mutual authentication.
Without this term, the protocol would only achieve unilateral authentication of the security module to the ECU and so would allow the adversary to impersonate any ECU.


\minisection{Stream Authorization}
The model of the stream authorization protocol described in \Cref{sec:Design_StreamAuthorization}.

While in reality, streams can be received by more than one \gls{ECU}, it is sufficient to verify the protocol with two participants, as this covers all messages transmitted.
In the case of an additional receiver, a duplicate message is added in the system, which does not affect the security.

The security module communicates with each ECU using the unique ECU key, \texttt{kesym}, established in the ECU authentication protocol.
In this model, we assume that all ECU keys have been securely established and use Scyther's built-in pairwise keys (e.g. \texttt{k(E1,SM)}) to represent ECU keys.
This is required to simplify the protocol representation and separately verify the \gls{ECU} authentication and stream authorization.
Having access to the stream key implicitly authorizes an ECU to participate in that stream and no further authentication is performed in order to minimize the communication latency.

All authorization decisions are made by the security module but these are orthogonal to the security claims of the communication protocol and are therefore beyond the scope of this model (see \Cref{sec:Scen_SystemSetup}).
The access to a stream is defined at design time by the system designer and is programmed into the security module, together with the certificates.
In this model, neither the sending nor receiving ECUs make security claims because they do not have control or visibility over which other \glspl{ECU} receive the stream key.
The stream key is distributed by the security model and can thus not be assumed as secret by a single \gls{ECU}.
These ECUs, therefore trust the security module to distribute the stream key correctly.
The influence of this on the security of the protocol is discussed in \Cref{sec:SecurityComparison}

The security module, which has overall visibility of the system, makes the following claims:

\begin{enumerate}

\item Secrecy of stream key: If none of the ECU keys have been compromised, then the newly established stream key cannot be known by the adversary. As the stream key is transmitted in messages encrypted with the \gls{ECU} key, this sets up a chain between \gls{ECU} and stream key.

\item Non-injective synchronization: On completion of a protocol run, all entities agree on the roles that have been taken and the data items that have been exchanged. Just as in the \gls{ECU} authentication, this synchronization is \emph{non-injective} as replays could be detected based on the message nonces and timestamps.

\end{enumerate}

We modeled the protocol in the proprietary language required by Scyther and defined the claims above.
Feeding this input to Scyther, it confirms that both of these claims are successfully verified in the unbounded sense.
This means that under the Dolev-Yao adversary model, this protocol ensures that the stream key is only known to the set of ECUs determined by the security module's authorization policy, thus authorizing these ECUs to participate in the message stream.

In this section, we successfully verified the security of the protocols in the lightweight authentication framework.
With the security being proven by a model checker, we now advance to evaluate the performance of the lightweight authentication framework with a discrete event simulation.
\section{Evaluation}
\label{sec:Evaluation}


In this section, we analyze and evaluate the characteristics and performance of our proposed approach.
For our evaluation, we developed a discrete event simulation of vehicular networks implementing our protocol (\cite{mundhenk2016SR}).
We use this simulation to evaluate the performance of the lightweight authentication framework by implementing a set of synthetic test cases.

\begin{figure}
	\centering
	\includegraphics[width=\hsize]{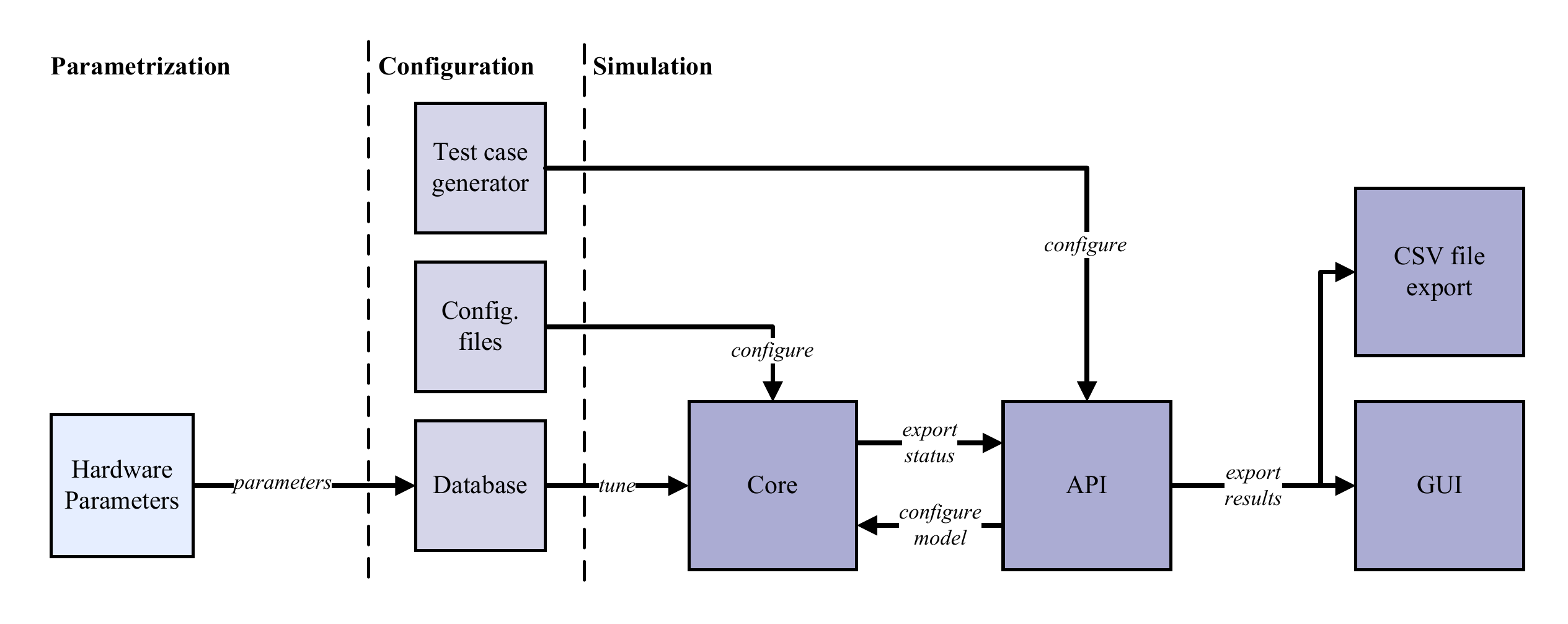}
	\caption{Structure of the developed discrete event simulator.
	Based on a set of measured hardware parameters, the simulator is configured with parameters, configuration and a test case.
	All simulation happens in the core and is accessible via an API.
	Results can be plotted in a GUI or exported into a CSV file.}
	\label{fig:simulatorStructure}
\end{figure}

\subsection{Simulation}
The core of our evaluation forms a newly developed simulator, designed for the real-time performance analysis of security in automotive networks.
The simulator is based on the discrete event simulaton framework SimPy \cite{simpy} and implemented in Python.
We model automotive networks in a modular fashion, representing the different components existing in networks and \glspl{ECU}.
The main components including \glspl{ECU}, gateways, and buses can be interconnected freely in any architecure.
These components themselves are made up of sub components, following the ISO/OSI model.
One of the components is the application layer, which can be flexibly programmed, allowing integration of any functional behavior, following actual \gls{ECU} implementations.
This allows a functional verification of software, a software-in-the-loop test or case studies for underlying communication schemes.
In a similar fashion, secure communication layers are implemented, following the specifications of LASAN, \gls{TESLA} and \gls{TLS}.
Description of the full set of simulation features is beyond the scope of this article.
The discrete event simulation is available open source for free use, development and public scrutiny \cite{mundhenk2016SR}.

The structure of the simulator is shown in \Cref{fig:simulatorStructure}.
It consists of modular simulator core, which implements most of the functionality, including the components and their submodules described above.
Further, an \gls{API} is provided to configure and run the simulation, as well as for the extraction of results.
A \gls{GUI} is provided, allowing direct inspection, debugging and evaluation of a running simulation.
The test case generator is used to generate test architectures based on statistic settings.
These test architectures are used to configure the model, before the simulator is run.
Additionally, the results may be exported into \gls{CSV} files for further analysis.

Further, the basic settings of the simulator, such as the use of the (optional) \gls{GUI} and the output paths can be set in configuration files.
These settings also include other basic settings such as timeouts, fallbacks for insufficiently configured components, etc.

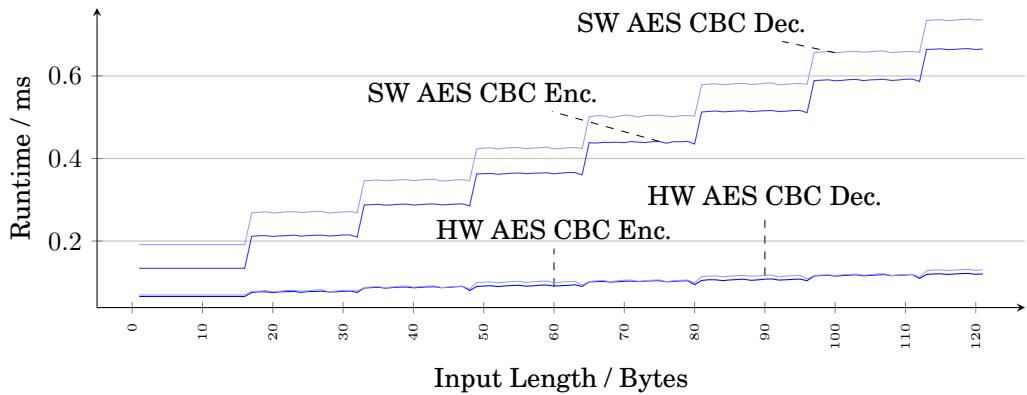
\begin{figure}
	\centering
\begin{tikzpicture}
\begin{axis}
[
	axis x line=bottom,
	axis y line=left,
	width=\textwidth,
	height=0.4\textwidth,
	x tick label style=
	{
		font=\tiny,rotate=90,anchor=east
	},
	legend pos=north west,
	xlabel=Input Length / Bytes,
	ylabel=Runtime / ms,
	enlarge y limits=0.04,
	enlarge x limits=0.05,
	extra x tick style=
	{
		grid=none,
		x tick label as interval=false
	},
	skip coords between index={121}{1536},
	ymajorgrids
]

\addplot+[mark=none, blue]
table[x=byte,y=HW_CBC_AVG_ENC]{results/STM32F415_HW_SW_comparison.txt};
\node[coordinate,pin={[pin distance=0.5cm,pin edge ={black, dashed}]+90:{HW AES CBC Enc.}}] at (axis cs:60,0.090875) {};

\addplot+[mark=none, blue!40]
table[x=byte,y=HW_CBC_AVG_DEC]{results/STM32F415_HW_SW_comparison.txt};
\node[coordinate,pin={[pin distance=0.8cm,pin edge ={black, dashed}]+90:{HW AES CBC Dec.}}] at (axis cs:90,0.1170625) {};

\addplot+[mark=none, blue!80]
table[x=byte,y=SW_CBC_AVG_ENC]{results/STM32F415_HW_SW_comparison.txt};
\node[coordinate,pin={[pin distance=0.8cm,pin edge ={black, dashed}]+150:{SW AES CBC Enc.}}] at (axis cs:75,0.441125) {};

\addplot+[mark=none, blue!40]
table[x=byte,y=SW_CBC_AVG_DEC]{results/STM32F415_HW_SW_comparison.txt};
\node[coordinate,pin={[pin distance=0.3cm,pin edge ={black, dashed}]+150:{SW AES CBC Dec.}}] at (axis cs:100,0.65625) {};
\end{axis}
\end{tikzpicture}
	\caption{
	AES performance for software implementations (SW) and with hardware support (HW).
	The measurements of encryptions (Enc.) and decryptions (Dec.) have been performed on an STM32 microcontroller with AES in Cipher Block Chaining (AES CBC) mode.
	Reproduced from \cite{mundhenk2015a}.
	}
	\label{fig:AES}
\end{figure}

\minisection{Timing Parameters}
To ensure that the simulator follows reality as close as possible, it can be parametrized with measurements from real hardware.
This parametrization is essential to ensure that correct results are achieved with the simulator.
All timings used in the simulator are derived from these basic timings.
By tuning the simulation according to hardware measurements, we are also able to simulate different hardware, simply by exchanging the timing parameters used.
These parameters vary significantly, based on the hardware and implementation chosen.
The computational capabilities and thus the speed for encryption/decryption on different \glspl{ECU} differ greatly.
Processor speed and hardware cryptographic support has a significant influence on timings.
This hardware support can come in the form of a \gls{TPM} or other hardware accelerators.
\Cref{tab:RSA} shows some example parameters for encryption/decryption latencies with RSA in a software implementation on an STM32 microcontroller.
Parameters for \gls{AES} with and without hardware support are shown in \Cref{fig:AES}.
Clearly visible are the block size of 128~Bit for \gls{AES} in \Cref{fig:AES}, as well as the very low performance of software implementations of RSA on microcontrollers, as shown in \Cref{tab:RSA}.
The full set of parameters includes many tens of thousands of measurements for different algorithms and configuration parameters and thus is exceeding what can be presented here.
All results are available online for free use \cite{ResultsRepo}.

\minisection{Test setup}
For our tests, we are interested in the setup process of message streams and thus do not require a functional verification of the data itself.
Therefore, we do not implement actual control algorithms, but transmit dummy data.
However, the selected message size and timing patterns follow existing distributions.

We base our communication on a \gls{CAN} \gls{FD} bus \cite{hartwich2012}.
\gls{CAN} \gls{FD} is an extension of \gls{CAN}, allowing to increase the data rate in the data portion of the frame, such that a higher overall throughput can be achieved.
Due to its low cost, similar to \gls{CAN}, and backwards compatibility, \gls{CAN} \gls{FD} is likely to find widespread use in future networks.
\gls{CAN} \gls{FD} is currently in the process of standardization.

The authentication frameworks presented here are implemented transparently to the application layer.
All handshakes and required messages are handled on the communication layers.
From the perspective of the application layer, the start time of the network is simply increased.

All results shown in this work have been generated with our simulator.
For the results displayed here, we built three different communication layers.
These include LASAN as proposed in this work, as well as the existing authentication frameworks \gls{TLS} and \gls{TESLA}.
\gls{TLS} is mostly used in Internet traffic, e.g. in HTTPS connections.
While not targeted for embedded systems or multicast communication, we implemented and used \gls{TLS} v1.2 as a baseline \cite{dierks2008}.
We further implemented a communication layer utilizing \gls{TESLA} \cite{perrig2005} to compare with the other approaches.
Both, \gls{TLS} and \gls{TESLA} have been implemented as timing correct.
That means that not all functionality is fully implemented, but the correct lengths of cryptographic algorithm runtimes, message lengths and transmission times are ensured.
On the other hand, to save memory and computational requirements for the simulator, identical certificates and keys are used for all \glspl{ECU}.
This is a reasonable approximation, as we are only interested in timing and not actual functionality of the protocols.

\begin{table}
\tbl{Latencies of RSA encryption and decryption in software for different RSA key lengths, measured on an STM32F415 microcontroller.}{
\centering
    \begin{tabular}{c|cc}
	    \toprule
	    \textbf{key length} & \textbf{public key}  & \textbf{private key}\\
	    (Bits) & \textbf{operations} & \textbf{operations}\\
	    \midrule
	    512 & 0.206s & 0.886s\\
	    1024 & 0.709s & 4.977s\\
	    2048 & 2.626s & 33.181s\\
	    \bottomrule
    \end{tabular}}
\label{tab:RSA}
\end{table}

\subsection{Security Comparison}
\label{sec:SecurityComparison}
Comparing the security of different approaches is no trivial task.
Security flaws are often found based on implementation flaws, not conceptual flaws.
As we are comparing the conceptual approaches, we keep the security comparison at a high level.
In the following, we will shortly detail the cryptographic algorithms used in the different approaches and explain differences in the security.

To level the playing field and allow equal conditions for all compared frameworks, we use the same cryptographic methods for all approaches.
For asymmetric functions, we use RSA with a key length of 512~Bits and an exponent of 65537.
We use MD5 for hashing algorithm for the certificate calculations.
For symmetric encryption, we use 128~Bit \gls{AES} in the \gls{CBC} mode.

We further assume that all keys used in the system have a limited security lifetime and must be renewed regularly.
This is typically the case, as long key lifetimes increase the chances of reverse engineering via brute force attacks.
It is necessary to exchange a key often enough that it is unlikely that a key can be brute forced within the given time.
Note that multiple message transmissions with the same key do not have any influence on the security of the key, but the potential insecurity is a result of a key's long validity.
Further, this time for brute force attacks decreases over time, as computational power increases, following Moore's Law.

Note that the \gls{TESLA} \gls{RFC} in \cite{perrig2005} defines the message security mechanism to be used as \gls{MAC}.
The \gls{RFC} does not consider encryption of messages.
While this is sufficient to protect the integrity of messages (i.e. no attacker can alter or create a message without detection), the confidentiality (i.e. an attacker can read a message) remains unprotected.
From the technical perspective, \gls{TESLA} could also use exchanged keys for encryption, but the \gls{RFC} only defines \glspl{MAC} and a strict implementation would need to follow this.
While this does not impact the security of the authentication and authorization mechanisms, it does heavily impact the security of data messages.

\gls{TESLA} further relies on a master key, securely exchanged before the conversation.
This exchange is not further defined in the \gls{RFC} \cite{perrig2005} and it is suggested to use existing mechanisms.
We assume here, similar to the other mechanisms, an initial key exchange via an asymmetric encryption scheme.

Further, note that \gls{TLS} only supports unicast messages.
This increases the security, as no keys are shared among any of the participants, but also increases setup time, as a separate key needs to be exchanged with every receiver, as well as bus load, as duplicate messages need to be sent for every receiver.

With LASAN, the symmetric message stream key is shared among all senders and receivers.
As symmetric cryptography is used to protect message transmissions, it would be possible for a malicious receiver that has circumvented the \gls{ECU} authentication and stream authorization to pretend to be the sender of a message stream and send message to all participants.
This is however, easily detectable on the approved sender side, as a message that should never have existed on the bus is received.
In this case, the current stream key can be voided and the communication participants (and possibly the user) can be notified of a breach in security.
It would be possible to enhance LASAN with either a reverse key chain and include a \gls{MAC}, such as in \gls{TESLA} or use asymmetric signatures to enforce directionality in the message transmission.
However, as we will show in \Cref{sec:LatencyComparison}, these approaches lead to significantly longer and unacceptable message transmission times.
As this illicit behavior is easy to detect on the sender side, we do not require these additional measures.

In summary, we can say that while there are slight differences in the concepts of the different approaches, the security of the authentication and authorization protocols should be comparable.

\subsection{Latency Comparison}
\label{sec:LatencyComparison}

\begin{figure}
	\centering
\begin{tikzpicture}
\begin{axis}
[
	cycle list = {black,black!40,black!60,black!80},
	axis x line=bottom,
	axis y line=left,
	width=0.78\textwidth,
	height=0.6\textwidth,
	x tick label style=
	{
		rotate=90,anchor=east
	},
	legend style={at={(0,0)}, anchor=north west,at={(axis cs:110,6000)}},
	xlabel=number of ECUs,
	ylabel=Start-up time for all streams / s,
	enlarge y limits=0.04,
	enlarge x limits=0.05,
	extra x tick style=
	{
		grid=none,
		x tick label as interval=false
	},
	ymajorgrids
]

\addplot+[mark=x, color=blue!40, only marks]
table[x=ecus,y=time]{results/summary/lw_auth/Crypto_Lib_HW/False/ECUAuth_raw.txt};
\addlegendentry{LASAN HW};

\addplot+[mark=x, color=blue, only marks]
table[x=ecus,y=time]{results/summary/lw_auth/CyaSSL/False/ECUAuth_raw.txt};
\addlegendentry{LASAN SW};

\addplot+[mark=o, color=blue!20, only marks]
table[x=ecus,y=time]{results/summary/tesla/Crypto_Lib_HW/False/Auth_raw.txt};
\addlegendentry{TESLA HW};

\addplot+[mark=o, color=blue!80, only marks]
table[x=ecus,y=time]{results/summary/tesla/CyaSSL/False/Auth_raw.txt};
\addlegendentry{TESLA SW};

\addplot[mark=+, color=black!40, only marks]
table[x=ecus,y=time]{results/summary/tls/Crypto_Lib_HW/False/Auth_raw.txt};
\addlegendentry{TLS HW};

\addplot[mark=+, color=black!80, only marks]
table[x=ecus,y=time]{results/summary/tls/CyaSSL/False/Auth_raw.txt};
\addlegendentry{TLS SW};

\coordinate (left1) at (axis cs:-6,7200);
\coordinate (right1) at (axis cs:105,7200);
\coordinate (right2) at (axis cs:108,7200);

\end{axis}
\draw [dashed] (left1) -- (right1);
\node[right] at (right2){timeout};
\end{tikzpicture}
	\caption{
		Comparison of start-up times in vehicles with varying numbers of ECUs and with different authentication frameworks (LASAN, TLS, TESLA) in software (SW) and hardware supported (HW) implementations.
		Architectures under test have been automatically generated and per number of ECUs, 5 test cases have been executed.
		Note that LASAN HW and SW are superimposed due to scale.
	}
	\label{fig:AuthRaw}
\end{figure}
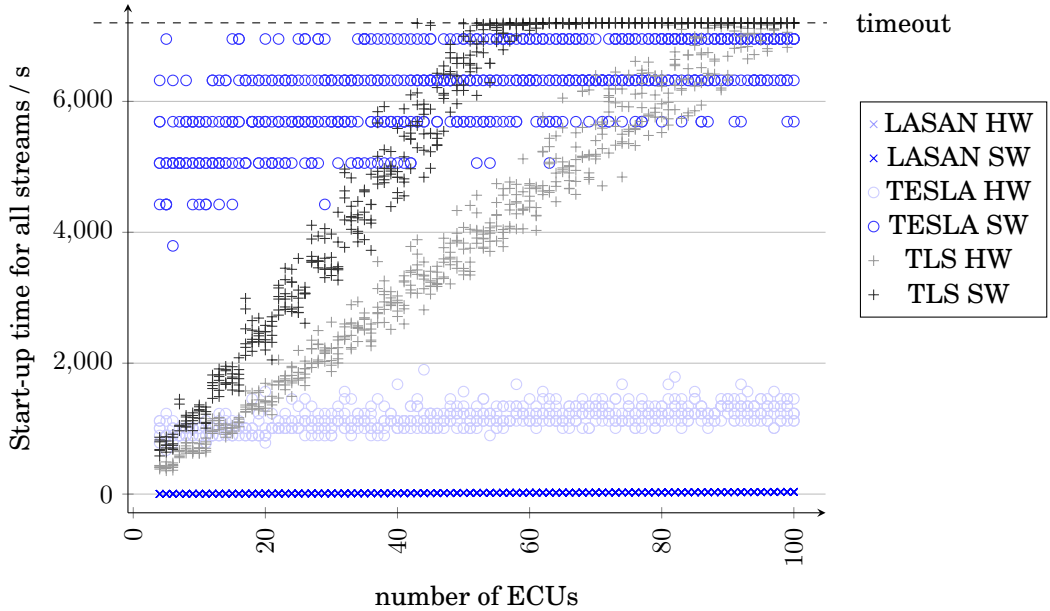

To evaluate the lightweight authentication framework, we use sets of randomly generated testcases from our simulator.
Based on the number of \glspl{ECU} in the system, an architecture is generated.
In this architecture, \glspl{ECU} are assigned randomly to buses, if more than one bus is used, a central gateway is added.
Messages are assigned random lengths and periods, based on real-world distributions and are distributed to the sending \glspl{ECU}.
To approach realistic network sizes, for every \gls{ECU} in the network, five messages are assumed.
Receiving \glspl{ECU} are randomly assigned.
Assignments of sending and receiving \glspl{ECU} are controlled by setting a guiding \gls{MAD} for the stream to \gls{ECU} distribution.

We compare LASAN with the existing authentication frameworks \gls{TLS} and \gls{TESLA}.
The results for a vehicle start-up without any previous authentication or authorization of any component or message are shown in Figure~\ref{fig:AuthRaw}.
There, we compare the worst-case startup-time, i.e. the time until all \glspl{ECU} are authenticated and all messages are requested and authorized at the same time, for different numbers of \glspl{ECU} and messages in the system.
Both hardware- and software-supported implementations of the authentication frameworks are shown.
In the following, we evaluate and interpret these results.

\minisection{TLS}
Clearly visible is the rapid latency increase for \gls{TLS} with the number of \glspl{ECU} and hence messages.
As in \gls{TLS} there is no common root of trust for message streams in the network, every message needs separate authentication between the sending \gls{ECU} and all receiving \glspl{ECU}.
The overall start-up time is thus defined by the \gls{ECU} sending or receiving most streams and thus requiring the longest time for encryption and decryption operations.
Even at medium size networks of 40 \glspl{ECU} onwards, the time for a full vehicle setup is in the range of one hour.
For slightly larger networks of around 50 \glspl{ECU}, start-up time reaches the simulation timeout of two hours.

\minisection{TESLA}
For \gls{TESLA}, the structure of the protocol is visible in the form of delays in the key generation.
As \gls{TESLA} requires keys to be generated before transmissions, the number of key sets to be generated depends on the number of streams to be sent by one \gls{ECU}.
The slowest \gls{ECU}, i.e. the \gls{ECU} with the most streams to send, defines the start-up time of the vehicle.
This key generation is visible as steps in Figure~\ref{fig:AuthRaw}.
In our testcases the number of messages sent by every \gls{ECU} is kept relatively constant, based on a given \gls{MAD} of 0.2.
This is represented by the limited number of steps in Figure~\ref{fig:AuthRaw}.
The key generation needs to be performed periodically.
In this setup, we are using 400,000 keys.
Once these keys are exhausted, a new set needs to be generated.
The period for key renewal is dependent on the period of the message, as for every transmission a single key is used.
A set of 400,000 keys and a message period of 10ms requires a new set of keys to be generated after about 1h.
With the number of \glspl{ECU} and message streams in the network increasing, more often than not, a single \gls{ECU} sends a larger set of messages.

Figure~\ref{fig:AuthRaw} clearly shows that hardware accelerated implementations of all authentication framework are significantly faster due to the faster encryption (compare Figure~\ref{fig:AES}).

It is also clearly visible that \gls{TESLA} and \gls{TLS} have not been designed for the automotive environment, requiring real-time constraints and support for messages with multiple receivers.
The key authentication (\gls{TLS}) and preparation (\gls{TESLA}) are far too large to be employed efficiently.

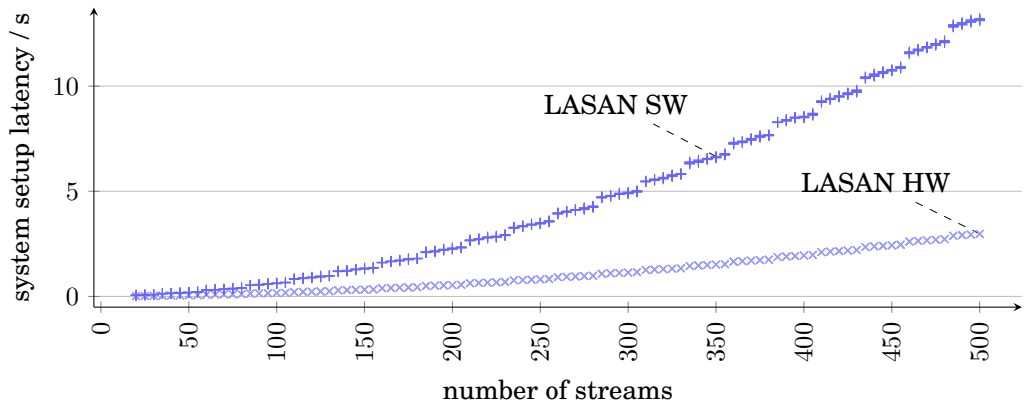
\begin{figure}
	\centering
\begin{tikzpicture}
\begin{axis}
[
	axis x line=bottom,
	axis y line=left,
	width=\textwidth,
	height=0.4\textwidth,
	x tick label style=
	{
		rotate=90,anchor=east
	},
	legend pos=north west,
	xlabel=number of streams,
	ylabel=system setup latency / s,
	enlarge y limits=0.04,
	enlarge x limits=0.05,
	extra x tick style=
	{
		grid=none,
		x tick label as interval=false
	},
	ymajorgrids
]

\addplot[mark=x, only marks, blue!40]
table[x=streams,y=time]{results/summary/lw_auth/Crypto_Lib_HW/True/StreamAuth_raw.txt};
\node[coordinate,pin={[pin distance=0.5cm,pin edge ={black, dashed}]+120:{LASAN SW}}] at (axis cs:350,6.654101060000348) {};

\addplot[mark=+, only marks, blue!60]
table[x=streams,y=time]{results/summary/lw_auth/CyaSSL/True/StreamAuth_raw.txt};
\node[coordinate,pin={[pin distance=0.5cm,pin edge ={black, dashed}]+122:{LASAN HW}}] at (axis cs:500,2.9765976100013485) {};
\end{axis}
\end{tikzpicture}
	\caption{
		Comparison of setup latencies for stream authorization in LASAN with software (SW) and hardware supported (HW) implementations over differing numbers of message streams.
	}
	\label{fig:StreamAuthResults}
\end{figure}

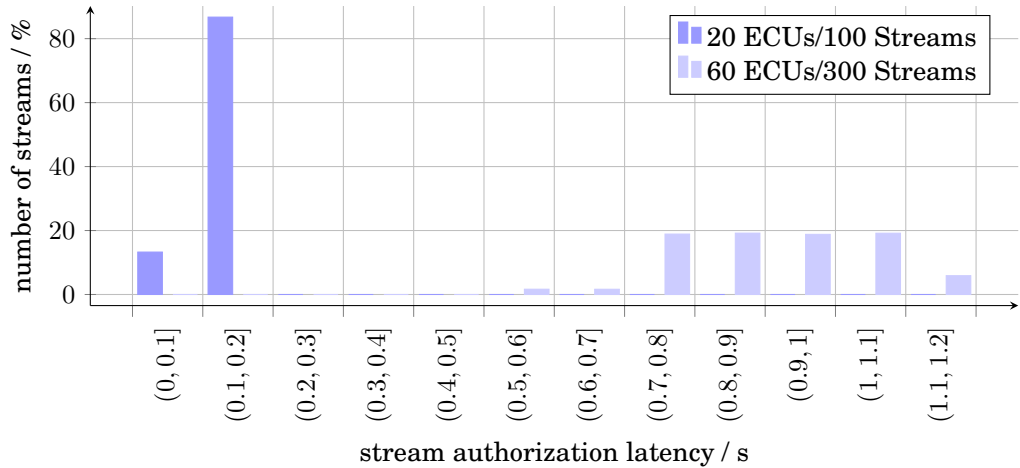
\begin{figure}
	\centering
\begin{tikzpicture}
\begin{axis}
[
	axis x line=bottom,
	axis y line=left,
	width=\textwidth,
	height=0.4\textwidth,
	x tick label style=
	{
		rotate=90,anchor=east
	},
	legend pos=north east,
	xlabel=stream authorization latency / s,
	ylabel=number of streams / \%,
	enlarge y limits=0.04,
	enlarge x limits=0.05,
	extra x tick style={
		grid=none,
		x tick label as interval=false,
		xticklabel=$\pgfmathprintnumber\tick$
	},
	xticklabel={$(\pgfmathparse{\tick-0.1}\pgfmathprintnumber[fixed]\pgfmathresult,\pgfmathprintnumber[fixed]\tick]$},
	ymajorgrids,
	ybar interval=0.7,
]

\addplot+[blue!40]
table[x=Bin,y=Frequency]{results/summary/lw_auth/Crypto_Lib_HW/True/100streams.txt};
\addlegendentry{20 ECUs/100 Streams};

\addplot+[blue!20]
table[x=Bin,y=Frequency]{results/summary/lw_auth/Crypto_Lib_HW/True/300streams.txt};
\addlegendentry{60 ECUs/300 Streams};

\end{axis}
\end{tikzpicture}
	\caption{
		Comparison of individual stream authorization times for architectures with 20 ECUs/100 Streams and 60 ECUs/300 Streams.
	}
	\label{fig:StreamAuthRelative}
\end{figure}


\minisection{LASAN}
Compared to \gls{TLS} and \gls{TESLA}, LASAN is significantly faster.
By minimizing the number of message transmissions and the sizes of messages in the authentication and authorization process, we reduced the start-up time to a minimum.
As shown in Figure~\ref{fig:AuthRaw}, LASAN is faster for all realistic network sizes, i.e. from 20 \glspl{ECU} onwards, even if only a software implementation is used.

LASAN is analyzed in more detail in Figure~\ref{fig:StreamAuthResults}.
There, a comparison of hardware and software implementations of the stream authorization is shown.
The setup latency as depicted here is the worst-case start-up time of the car, if all messages in the vehicle are authorized at the same time.
This scenario is rather unlikely, as event-based messages without real-time properties, such as button presses by the driver, only need to be authorized when transmitted, thus reducing the number of streams significantly.
Further, Figure~\ref{fig:StreamAuthResults} depicts the maximum setup latency of all streams in the vehicle.
By contrast, Figure~\ref{fig:StreamAuthRelative} shows the individual setup latencies of all streams in two architecture configurations.
It is to be noted that a large number of streams experience lower latencies than the overall system latency given in Figure~\ref{fig:StreamAuthResults}.
Thus, the stream setup could be prioritized, ensuring that streams required earlier are authorized earlier.
The scheduling of stream setups according to priorities is an optimization topic in itself and is out-of-scope of this work.

When analyzing Figure~\ref{fig:StreamAuthResults}, the exponential behavior is to be noted.
This results from the additionally required authorization messages including the symmetric stream key for additional receivers.
The resulting increase is acceptable even for large systems with 500 messages, as shown in Figure~\ref{fig:StreamAuthResults}.
Increases beyond this number of messages are unlikely in the automotive domain.

The clustering of messages can be explained through the architecture generation.
To model a realistic system, where messages are received and processed by multiple components, the number of receivers is increased by 1 every 25 streams.
As the additional receiver also needs a grant message with the stream key, the start-up time rises.


When comparing software and hardware supported implementations of LASAN, the advantage of hardware supported implementations can be clearly seen (see Figure~\ref{fig:StreamAuthResults}).
By utilizing the increasingly existing hardware accelerators in \glspl{ECU}, in a system with 500 message streams the start-up time can be reduced from 13.18s to 2.97s.
Table~\ref{tab:singleStream} compares the delays introduced by LASAN into a single stream setup.
If \gls{ECU} authentication is completed and hardware support is available, a stream can be set up in less than 2ms.
This is by far sufficient when comparing it with most conventional messages, such as in the electric test vehicle EVA \cite{eva}.
EVA is a fully functional electric taxi, built as a research platform by TUM CREATE, supporting research into electric vehicles.
In EVA, no message is transmitted with a period under 20ms and all messages are specified to tolerate a delay of up to 80ms.
Modern \glspl{ADAS} might have more strict requirements below 10ms, but even here, the setup time is still appropriate.
Additionally, for extremely high real-time constraints, a stream can be authorized at vehicle startup, reducing the setup delay required at the time of message transmission to zero.

\minisection{Future Architectures}
Future automotive architectures are expected to supply more bandwidth, computational power and memory, while at the same time reducing the overall number of \glspl{ECU}.
This \gls{ECU} consolidation will lead to LASAN gaining efficiency in the future.
The higher computational power and memory can support larger key lengths and stronger cryptographic measures while driving down the latencies LASAN is currently experiencing.
The higher bandwidth will further contribute to a reduction in stream setup latencies.
\gls{ECU} consolidation, also with the possible consolidation of message streams, would decrease vehicle start-up delays in LASAN, as delays are mostly dependent on the number of receivers and message streams.

\minisection{Summary}
As shown in the preceding evaluations, LASAN offers low latency stream authorization, while keeping \gls{ECU} authentication in acceptable boundaries.
Compared to existing authentication frameworks LASAN performs highly efficiently, lowering latencies in typical network sizes of 60 to 100 \glspl{ECU} by on average factor 49 for \gls{TESLA} and factor 234 for \gls{TLS}.
Yet, LASAN can be improved by scheduling stream authorization based on priority and only when required.
This way, stream authorization overhead can be reduced down to 1.6ms, a value acceptable for all but the applications with the strictest real-time requirements.
For these applications, streams can be set up a priori, before real-time responses are required.

\begin{table}
\tbl{Stream setup latencies using LASAN for a single stream in ideal conditions (2 ECUs, single bus, no cross traffic).}{
\centering
    \begin{tabular}{c|cc}
	    \toprule
	    \textbf{ECU Authentication} & \multicolumn{2}{c}{\textbf{Setup Latency}}\\
	    & \textbf{Hardware} & \textbf{Software}\\
	    \midrule
	    included & 1.4s & 2.3s\\
	    not included & 0.0016s & 0.0046s\\
	    \bottomrule
    \end{tabular}}
\label{tab:singleStream}
\end{table}

\section{Concluding Remarks}
\label{sec:Conclusion}
In this work we present the design, evaluation, verification and integration of the Lightweight Authentication for Secure Automotive Networks (LASAN).
We designed LASAN to achieve high performance, even in environments with low computational power and network bandwidth.
We showed that the protocols contained in LASAN can be verified with standard network protocol verification tools and thus are proven to be secure to known attacks.
We evaluated the performance of LASAN with a purposely developed discrete event simulation and showed latency decreases of factor 49 to 234 for typical network sizes over existing frameworks, while keeping the security steady.
This has been achieved through reducing message sizes, reducing the number of messages, and optimizing use of protocol components for automotive environments.
An acceptable loss of runtime flexibility in static automotive networks is compensated for by a secure and real-time capable authentication and authorization of all network participants and messages.
As with any digital security measure, the security strength of LASAN is highly dependent on the key length used.
With rising computational power and memory, LASAN can make use of greater key lengths, thereby maintaining security.
LASAN has further been shown to integrate well with and support existing scenarios in the automotive domain, allowing and supporting secure firmware and \gls{ECU} replacements and updates from the production of the vehicle till the end-of-life.
To the best of our knowledge this is a first for an automotive authentication mechanism.

While this work focused on the evaluation of LASAN, future work will be concerned with optimizing the framework based on the results acquired in this evaluation.
By scheduling \gls{ECU} authentication and stream authorization messages in a priority-based fashion, the performance for individual messages and thus components and control systems can be increased significantly.
This can be achieved in the same fashion as optimization for safety-critical applications, while accounting for the security specifics.
Further, the security of LASAN, as for every security measure, heavily depends on the correctness of the implementation.
To guarantee this, a reference implementation of LASAN and the discrete event simulation used for evaluation is planned to be published as open source.
Further, the basic verification of LASAN, as shown in this work, shall be extended to allow verification of real-world LASAN implementations (Hardware-in-the-Loop).

\bibliographystyle{ACM-Reference-Format-Journals}
\bibliography{library}


\end{document}